\documentclass[iop]{emulateapj}

\pdfoutput=1
\newcommand\xone{x_1}
\newcommand\xtwo{x_2}
\newcommand\R{{\mathcal R}}
\newcommand\C{{\mathcal C}}
\newcommand\vyone{\dot{y}_{x_1}}
\newcommand\vytwo{\dot{y}_{x_2}}
\newcommand\uy{u_y}
\newcommand\Qb{Q_b}
\newcommand\Qm{Q_m}
\newcommand\fbar{f_{\rm bar}}
\newcommand\RQ{r(\Qb)}
\newcommand\Mtot{M_{\rm tot}}
\newcommand\Mbar{M_{\rm bar}}
\newcommand\Mbul{M_{\rm bul}}
\newcommand\Dalp{\Delta\alpha}
\newcommand\cs{c_s}
\newcommand\Pbar{\Phi_{\rm bar}}
\newcommand\Ptot{\Phi_{\rm ext}}

\newcommand\Msun{\; {\rm M}_{\odot}}
\newcommand\kms{\; {\rm km}\;{\rm s}^{-1}}
\newcommand\kpc{\;{\rm kpc}}
\newcommand\freq{\kms\kpc^{-1}}

\newcommand\yr{\; {\rm yr}}

\newcommand\Gyr{\;{\rm Gyr}}

\newcommand\Surf{\Msun\;{\rm pc^{-2}}}
\newcommand\MBH{M_{\rm BH}}
\newcommand\Rmin{r_{\rm min}}
\newcommand\Rmax{r_{\rm max}}
\newcommand\RCR{r_{\rm CR}}
\newcommand\ILR{r_{\rm ILR}}

\newcommand\Omb{\Omega_{b}}
\newcommand\rhobar{\rho_{\rm bar}}

\newcommand\ip{i_p}
\newcommand\xp{x_{\rm peak}}
\newcommand\pSig{\Sigma_{\rm peak}}
\newcommand\Sring{\Sigma_{\rm ring}}
\newcommand\Rring{r_{\rm ring}}
\newcommand\ering{\epsilon_{\rm ring}}
\newcommand\pring{\phi_{\rm ring}}
\newcommand\Mdot{\dot{M}}
\newcommand\Aunit{\Msun\yr^{-1}}
\newcommand\simgt{\lower.5ex\hbox{$\; \buildrel > \over \sim \;$}}
\newcommand\simlt{\lower.5ex\hbox{$\; \buildrel < \over \sim \;$}}

\shorttitle{Gaseous Structures in Barred Galaxies}
\shortauthors{Kim et al.}

\begin{document}

\title{Gaseous Structures in Barred Galaxies: Effects of the Bar Strength}
\author{Woong-Tae Kim, Woo-Young Seo, \& Yonghwi Kim}
\affil{Center for the Exploration of the Origin of the Universe
(CEOU), Astronomy Program, Department of Physics \& Astronomy, Seoul
National University, Seoul 151-742, Republic of Korea} \affil{FPRD,
Department of Physics \& Astronomy, Seoul National University, Seoul
151-742, Republic of Korea} \email{wkim@astro.snu.ac.kr}
\slugcomment{Accepted for publication in the ApJ}

\begin{abstract}
Using hydrodynamic simulations, we investigate the physical
properties of gaseous substructures in barred galaxies and their
relationships with the bar strength. The gaseous medium is assumed
to be isothermal and unmagnetized. The bar potential is modeled as a
Ferrers prolate with index $n$. To explore situations with differing
bar strength, we vary the bar mass $\fbar$ relative to the
spheroidal component as well as its aspect ratio $\R$. We derive
expressions as functions of $\fbar$ and $\R$ for the bar strength
$\Qb$ and the radius $\RQ$ where the maximum bar torque occurs. When
applied to observations, these expressions suggest that bars in real
galaxies are most likely to have $\fbar\sim0.25$--$0.50$ and
$n\simlt1$. Dust lanes approximately follow one of $\xone$-orbits
and tend to be more straight under a stronger and more elongated
bar, but are insensitive to the presence of self-gravity.  A nuclear
ring of a conventional $\xtwo$ type forms only when the bar is not
so massive or elongated. The radius of an $\xtwo$-type ring is
generally smaller than the inner Lindblad resonance, decreases
systematically  with increasing $\Qb$, and slightly larger when
self-gravity is included. This evidences that the ring position is
not determined by the resonance but by the amount of angular
momentum loss at dust-lane shocks. Nuclear spirals exist only when
the ring is of the $\xtwo$-type and sufficiently large in size.
Unlike the other features, nuclear spirals are transient in that
they start out as being tightly-wound and weak, and then due to the
nonlinear effect unwind and become stronger until turning into
shocks, with an unwinding rate higher for larger $\Qb$. The mass
inflow rate to the galaxy center is found to be less than
$0.01\Aunit$ for models with $\Qb\simlt0.2$, while becoming larger
than $0.1\Aunit$ when $\Qb\simgt0.2$ and self-gravity is included.
\end{abstract}

\section{Introduction\label{sec:intro}}

The inner parts of barred galaxies contain interesting gaseous
substructures that include dust lanes, nuclear rings, and nuclear
spirals (e.g., \citealt{san76,sak99,kna02,
mar03a,mar03b,she05,pri05,mar06,com09,com10,maz11,hsi11}). These
substructures are thought to form as a result of gas redistribution
in galaxies initiated by non-axisymmetric bar torque (e.g.,
\citealt{com85,but86,shl90,gar91,but96,com01}). Understanding their
formation and evolution is therefore of crucial importance to
understand how the interstellar gas in barred galaxies is driven
inward to affect star formation activities in the nuclear regions
and potentially fueling of active galactic nuclei (AGN). Since their
spatial locations and shapes are likely to be determined by the bar
strength as well as the underlying rotation curve (e.g.,
\citealt{ath92b,pee06,com09,com10}), they may also be used to probe
the mass distributions in barred galaxies that are not directly
observable.

While the bar substructures have been observed for a long time
(e.g., \citealt{pea17,san61}), it is during recent years that
high-quality data on their physical properties have been compiled
and compared with bar characteristics (e.g.,
\citealt{kna02,pee06,maz08,com09,com10,maz11}). There is increasing
observational evidence that the shape and size of the bar
substructures are determined primarily by the bar strength $\Qb$.
For instance, \citet{kna02} measured the curvature angles $\Dalp$ of
dust lanes for a sample of 9 barred-spiral galaxies, and found that
stronger bars favor more straight dust lanes, confirming a
theoretical prediction of \citet{ath92b}.  By extending the sample
size to 55 galaxies, \citet{com09} confirmed the results of
\citet{kna02}, although they also found that a large scatter in the
fit of the $\Dalp$--$\Qb$ relationship can be reduced if the bar
ellipticity is also considered in the fit.  More recently,
\citet{com10} measured the sizes and shapes of nuclear rings in a
sample of 107 galaxies, finding that `stronger bars host smaller
rings' and that the ring ellipticity is in the range of 0--0.4.
\citet{maz11} reported from the analysis of 13 nuclear rings that
the ring size is well correlated with the compactness of the galaxy
mass distribution such that higher mass concentration implies a
smaller ring. On the other hand, by analyzing dust morphology in the
nuclear regions of 75 galaxies, \citet{pee06} found that
tightly-would nuclear spirals are preferentially found in galaxies
with a weak bar.

On the theoretical side, the formation and evolution of bar
substructure has been extensively studied using numerical
simulations on hydrodynamical models (e.g.
\citealt{san76,rob79,van81,ath92b,pin95,eng97,pat00,mac02,mac04a,
mac04b,reg03,reg04,ann05,lin08,tha09}) and magnetized models (e.g.,
\citealt{kul09,kul10,kul11,ks12}). In particular, \citet{ath92b}
confirmed the early notion of \citet{pre62} that dust lanes are
shocks in the gas flows. She also showed that dust lanes are more
straight when the bar potential is stronger.  Due to lack of
numerical resolution and/or large numerical viscosity, however,
early numerical works published before the middle of 1990s were
unable to capture the formation of nuclear rings and spirals
accurately. \citet{pin95} was the first who focused on the ring
formation using a grid-based code with negligible numerical
viscosity. Unfortunately, however, their numerical results were, at
least quantitatively, contaminated by a trivial sign error in the
evaluation of the bar forces, as identified by \citet{kim12}. More
recently, \citet{ann05} and \citet{tha09} used SPH simulations to
study gas dynamics in the central regions of barred galaxies, but
they were unable to resolve nuclear spirals because most particles
in their simulations are captured into a ring, leaving only a small
number of particles inside the ring.

Very recently, \citet[hereafter Paper I]{kim12} corrected the error
in \citet{pin95} and revisited the issue of the bar-substructure
formation using high-resolution simulations by varying the gas sound
speed $\cs$ and the mass of a central black hole (BH). Paper I found
that the corrected bar force naturally allows the development of
nuclear spirals inside a ring, which was unseen in \citet{pin95}. In
addition, Paper I showed that the ring size and the mass inflow rate
$\Mdot$ toward the galaxy center depend rather sensitively on $\cs$.
In models with small $\cs$, the rings are narrow and located away
from the center. This prevents further inflows of the gas to the
central regions, making $\Mdot$ smaller than $10^{-4}\Aunit$. On the
the other hand, models with large $\cs$ have a small and broad ring,
resulting in $\Mdot\simgt10^{-2}\Aunit$. Paper I further confirmed
that the prediction of \citet{but96} that the shape of nuclear
spirals depend on the sign of the $d(\Omega-\kappa/2)/dr$ curve such
that they are trailing (leading) where $d(\Omega-\kappa/2)/dr$ is
negative (positive). This suggests that nuclear spirals, if exist,
are likely to be trailing in galaxies with supermassive BHs.

Since the models studied in Paper I were restricted to a particular
set of the bar parameters, namely, with the bar mass $\sim30\%$ of
the spheroidal component (i.e., bulge plus bar) and the aspect ratio
$\R=2.5$, they were unable to study the variations in the size and
shape of bar substructures with differing bar strength.  In
addition, the models in Paper I are all non-self-gravitating, so
that the effects of self-gravity on bar substructures have yet to be
explored. Although \citet{ath92b} considered diverse bar models with
different aspect ratio and bar quadrupole moment $\Qm$, her models
were unable to resolve nuclear rings and spirals due to insufficient
resolution. While \citet{reg03,reg04} also ran a large number of
numerical simulations with varying $\Qm$ and $\R$, their numerical
results were compromised by the incorrect bar forces of
\citet{pin95}, as mentioned above.

In this paper, we present the results of a series of hydrodynamic
simulations to investigate the properties of bar substructures that
form. This work extends Paper I in two ways: (1) by considering
various bar models with differing mass and aspect ratio and (2) by
including gaseous self-gravity, while fixing the sound speed and the
BH mass. We measure the curvatures of dust lanes, sizes of nuclear
ring, shapes of nuclear spirals, and mass inflow rates in
simulations, and study their relationships with the bar parameters.
Our main objective is to explore the parametric dependence of the
properties of bar substructures on the bar strength and
self-gravity. This will allow us to provide physical explanations
for the formation of bar substructures especially for nuclear rings
and spirals, which was previously uncertain. We also compare our
numerical results with observations of barred-spiral galaxies that
are currently available.

Our models with varying bar strength are useful to address an
important question as to what determines the position of nuclear
rings. Observations indicates that nuclear rings are typically
located near the inner Lindblad resonance (ILR) when there is a
single ILR, or between the ILRs when there are two ILRs (e.g.,
\citealt{com85, kna95,com10}). This has often been interpreted as an
indication that nuclear rings form as a consequence of the resonant
interactions of the gas with the ILRs (\citealt{com96,but96}; see
also \citealt{reg03}). This idea was motivated by the notion that
the bar torque changes its sign whenever crossing each ILR such that
in the case of a single ILR, for example,  the gas inside (outside)
the ILR receives a positive (negative) torque and thus moves outward
(inward). This idea of the resonance-driven ring formation requires
that the bar torque dominates thermal and ram pressures of the gas
in governing the gas motions both inside and outside the ILR.  It is
clear that the bar torque dominates outside the ILR, inducing
dust-lane shocks and initiating radial gas inflows. However, it is
uncertain if that is also the case inside the ILR.  The bar
potential becomes increasingly axisymmetric toward the galaxy
center, while the inflowing gas usually has large momentum and is
thus unlikely to stall at the ILR. By measuring the ring positions
and comparing them with the ILR radii in our models, we directly
test whether the ILR is really responsible for the formation of
nuclear rings.

This paper is organized as follows: In Section 2, we describe our
model parameters and numerical methods.  In Section 3, we define and
evaluate the bar strength $\Qb$ of our galaxy models as well as the
radius $\RQ$ where the maximum bar torque occurs.  We provide
algebraic expressions for $\Qb$ and $\RQ$ for various bar models. In
Section 4, we quantify the properties of bar substructures and
explore their correlations with the bar strength. We also measure
the mass inflows rates. Finally, in Section 5, we summarize our
results and discuss them in comparison with observations.

\section{Model\label{sec:model}}

We study gas responses to an imposed non-axisymmetric bar potential
using hydrodynamic simulations with and without self-gravity,
focusing on the effects of the bar strength on gaseous structures
that form. We consider an initially-uniform, rotating gaseous disk
with surface density $\Sigma_0=10\Surf$.  We assume that the disk is
infinitesimally thin, isothermal, and unmagnetized. The simulations
setups and numerical methods are identical to those in Paper I,
except that some models in the present work include
self-gravitational potential of the gas. Here we briefly summarize
our numerical models and highlight the differences between the
current models and those studied in Paper I.

\begin{deluxetable*}{lcccccccc}
\tabletypesize{\footnotesize} \tablewidth{0pt} \tablecaption{Model
Parameters and Curvatures of Dust Lanes in Non-self-gravitating
Models \label{tbl:model_no}} \tablehead{ \colhead{Model}   &
\colhead{$\fbar$} & \colhead{$\R$} & \colhead{$\ILR$}   &
\colhead{$\Qb$}   & \colhead{$\RQ$} & \colhead{$\alpha_1$} &
\colhead{$\alpha_2$} & \colhead{$\Dalp$}
\\
\colhead{     } & \colhead{     } & \colhead{     } &
\colhead{(kpc)} & \colhead{     } & \colhead{(kpc)} &
\colhead{(deg)} & \colhead{(deg)} & \colhead{(deg)}
\\
\colhead{(1)} & \colhead{(2)} & \colhead{(3)} & \colhead{(4)} &
\colhead{(5)} & \colhead{(6)} & \colhead{(7)} & \colhead{(8)} &
\colhead{(9)} } \startdata
M08R15N & 0.08  &1.5 & 2.2& 0.02 & 3.60 & 70.5 $\pm$ 0.5 & 35.8 $\pm$ 0.9 & 64.2 $\pm$  2.5 \\
M08R20N & 0.08  &2.0 & 2.2& 0.05 & 3.30 & 74.4 $\pm$ 5.8 & 35.8 $\pm$ 2.9 & 63.4 $\pm$ 12.6 \\
M08R25N & 0.08  &2.5 & 2.2& 0.07 & 3.12 & 74.1 $\pm$ 2.6 & 44.2 $\pm$ 2.3 & 46.3 $\pm$  3.4 \\
M08R30N & 0.08  &3.0 & 2.2& 0.09 & 3.04 & 72.8 $\pm$ 3.0 & 53.3 $\pm$ 4.0 & 30.1 $\pm$  6.6 \\
M08R35N & 0.08  &3.5 & 2.2& 0.12 & 2.98 & 74.3 $\pm$ 1.7 & 59.6 $\pm$ 5.2 & 22.2 $\pm$  6.7 \\
\hline
M15R15N & 0.15  &1.5 & 2.1& 0.04 & 3.58 & 75.1 $\pm$ 1.4 & 34.5 $\pm$ 1.6 & 81.6 $\pm$  2.9 \\
M15R20N & 0.15  &2.0 & 2.1& 0.08 & 3.28 & 71.7 $\pm$ 3.2 & 49.6 $\pm$ 4.3 & 36.3 $\pm$  9.0 \\
M15R25N & 0.15  &2.5 & 2.2& 0.13 & 3.12 & 75.3 $\pm$ 2.5 & 63.4 $\pm$ 2.3 & 19.6 $\pm$  4.2 \\
M15R30N & 0.15  &3.0 & 2.2& 0.17 & 3.04 & 73.4 $\pm$ 2.5 & 68.3 $\pm$ 2.4 &  8.2 $\pm$  4.4 \\
M15R35N & 0.15  &3.5 & 2.2& 0.21 & 2.98 & 74.3 $\pm$ 5.1 & 71.1 $\pm$ 5.4 &  5.0 $\pm$  3.6 \\
\hline
M30R15N & 0.30  &1.5 & 1.8& 0.08 & 3.56 & 70.7 $\pm$ 3.4 & 38.9 $\pm$ 3.9 & 63.4 $\pm$ 10.0 \\
M30R20N & 0.30  &2.0 & 1.8& 0.16 & 3.26 & 78.8 $\pm$ 3.1 & 60.6 $\pm$ 4.5 & 33.4 $\pm$  9.6 \\
M30R25N & 0.30  &2.5 & 2.1& 0.23 & 3.10 & 79.8 $\pm$ 1.5 & 68.5 $\pm$ 1.4 & 16.8 $\pm$  2.8 \\
M30R30N & 0.30  &3.0 & 2.1& 0.31 & 3.02 & 78.4 $\pm$ 2.8 & 75.1 $\pm$ 0.9 &  4.7 $\pm$  3.9 \\
M30R35N & 0.30  &3.5 & 2.1& 0.39 & 2.90 & 80.8 $\pm$ 0.3 & 79.3 $\pm$ 0.5 &  2.2 $\pm$  0.2 \\
\hline
M60R15N & 0.60  &1.5 & 1.4& 0.14 & 3.42 & 81.0 $\pm$ 3.3 & 52.6 $\pm$ 11.1& 53.4 $\pm$ 23.1 \\
M60R20N & 0.60  &2.0 & 1.3& 0.28 & 3.18 & 79.2 $\pm$ 3.1 & 70.7 $\pm$  1.1& 12.8 $\pm$  5.7 \\
M60R25N & 0.60  &2.5 & 1.3& 0.41 & 3.02 & 79.3 $\pm$ 3.4 & 74.8 $\pm$  1.9&  6.9 $\pm$  5.0 \\
M60R30N & 0.60  &3.0 & 1.9& 0.54 & 2.92 &      $\cdots$  &      $\cdots$  &      $\cdots$   \\
M60R35N & 0.60  &3.5 & 1.9& 0.67 & 2.88 &      $\cdots$  & $\cdots$
& $\cdots$
\enddata
\tablecomments{$\fbar$ is the fraction of the bar mass relative to
the spheroidal component (bar plus bulge); $\R$ is the ratio of the
bar semimajor axis to the semiminor axis; $\ILR$ is the radius of
the ILR; $\Qb$ is the bar strength defined by equation
(\ref{eq:Qb}); $\RQ$ is the radius of the maximum bar torque;
$\alpha_1$ and $\alpha_2$ are the tangent angles to a dust-lane
segment at the inner and outer ends relative to the $x$-axis,
respectively; $\Dalp$ is the dust lane curvature defined by equation
(\ref{eq:dalp}).}
\end{deluxetable*}

The gaseous disk is placed under the external gravitational
potential $\Ptot$ consisting of four components: a stellar disk, a
stellar bulge, a non-axisymmetric stellar bar, and a central BH. The
stellar disk is modeled by a Kuzmin-Toomre profile, the bulge by a
modified Hubble profile, and the BH by a Plummer profile with mass
$\MBH$. For the bar potential $\Pbar$, we use \citet{fer87} prolate
spheroids with volume density
\begin{eqnarray}\label{eq:bar}
\rho =
\left\{ \begin{array}{ll}
\rhobar \left( 1-g^2 \right)^n, & ~~~\textrm{for}~ g<1, \\
~~~~~~~0, & ~~~\textrm{elsewhere},
\end{array} \right.
\end{eqnarray}
where $\rhobar$ is the central density, $g^2\equiv y^2/a^2 +(x^2
+z^2)/b^2$, and $a$ and $b$ $(\leq a)$ denote the semimajor and
semiminor axes of the bar, respectively. The index $n$ represents
the degree of central density concentration. Appendix presents an
analytic expression of $\Pbar$ at $z=0$ and $g<1$ for $n=1$.

The total mass of the bar is given by $\Mbar = 2^{2n+3}\pi a
b^2\rhobar \Gamma(n+1)\Gamma(n+2)/\Gamma(2n+4)$ (e.g.,
\citealt{ath09}), which we control using a dimensionless parameter
\begin{equation}\label{eq:fbar}
\fbar = \frac{\Mbar}{\Mbar+\Mbul},
\end{equation}
where $\Mbul$ denotes the bulge mass inside $r=10\kpc$. In our
models, the total mass of the spheroidal component within $10\kpc$
is fixed to $\Mbar+\Mbul=4.87\times10^{10}\Msun$, so that larger
$\fbar$ implies a less massive bulge. Another independent parameter
that characterizes the bar potential is the aspect ratio $\R\equiv
a/b$. One may also use the bar quadrupole moment $\Qm =
\Mbar(a^2-b^2)/(5+2n)$ as a measure of the bar mass (e.g.,
\citealt{ath92b,reg03,reg04}), although we prefer $\fbar$ since
$\Qm$ depends also on $\R$. The rotation curve in the bar regions is
slightly changed by varying $\fbar$ and $\R$, but its flat part has
$v_0=200\kms$ regardless of $\fbar$ and $\R$, corresponding to
normal disk galaxies.

Since the parameter space is very large, Paper I fixed the bar
parameters to $\fbar=0.3$, $a=5\kpc$, and $b=2\kpc$, and varied
$\cs$ from 5 to $20\kms$ and $\MBH$ from 0 to $4\times 10^8 \Msun$
to explore how gas flows change with the gas sound speed and the BH
mass. In this work, we instead fix $\cs=10\kms$ and $\MBH=4\times
10^7 \Msun$, and consider 40 models with differing $\fbar=0.08$ to
0.6 and $\R=1.5$ to 3.5. Since the bar semimajor axis is fixed to
$a=5\kpc$ in all of our models, the latter is equivalent to varying
$b$ from $3.33$ to $1.43\kpc$. Other parameters such as the bar
pattern speed $\Omb=33\freq$ and the total disk mass $M_{\rm disk} =
2.2\times 10^{11}\Msun$ remain the same as in Paper I. Columns (1)
-- (3) of Tables \ref{tbl:model_no} and \ref{tbl:model_sg} list the
model names and parameters for non-self-gravitating and
self-gravitating models, respectively. The postfixes N and G stand
for no self-gravity and with self-gravity, respectively.  In what
follows, a model name without any postfix refers to both
self-gravitating and non-self-gravitating models. Note that Model
M30R25N is identical to Model cs10bh7 studied in Paper I except for
the size of the inner boundary (see below).

With $\Omb=33\freq$, all the models have the corotation resonance at
$\RCR=6\kpc$, independent of $\fbar$ and $\R$. The presence of a
central BH makes the rotation curve rise steeply toward the center
as $\propto(\MBH/r)^{1/2}$, giving rise to a single ILR where
$\Omb=\Omega-\kappa/2$. Here, $\Omega$ and $\kappa$ refer to the
angular and epicycle frequencies, respectively, averaged between on
the major and minor axes of the bar. Unlike $\RCR$, the ILR position
$\ILR$ depends slightly on $\fbar$ and $\R$. Column (4) of Tables
\ref{tbl:model_no} and \ref{tbl:model_sg} gives $\ILR$ in each
model. Since the bulge potential is more centrally concentrated than
the bar potential, models with larger $\fbar$ tend to have a smaller
rotational velocity in the central regions and hence smaller $\ILR$.
For fixed $\fbar$, a larger value of $\R$ corresponds to larger
$\rhobar$ and thus slightly larger $\ILR$.

\begin{deluxetable*}{lcccccccc}
\tabletypesize{\footnotesize} \tablewidth{0pt} \tablecaption{Model
Parameters and Curvatures of Dust Lanes in Self-gravitating Models
\label{tbl:model_sg}} \tablehead{ \colhead{Model}   &
\colhead{$\fbar$} & \colhead{$\R$} & \colhead{$\ILR$}   &
\colhead{$\Qb$}   & \colhead{$\RQ$} & \colhead{$\alpha_1$} &
\colhead{$\alpha_2$} & \colhead{$\Dalp$}
\\
\colhead{     } & \colhead{     } & \colhead{     } &
\colhead{(kpc)} & \colhead{     } & \colhead{(kpc)} &
\colhead{(deg)} & \colhead{(deg)} & \colhead{(deg)}
\\
\colhead{(1)} & \colhead{(2)} & \colhead{(3)} & \colhead{(4)} &
\colhead{(5)} & \colhead{(6)} & \colhead{(7)} & \colhead{(8)} &
\colhead{(9)} } \startdata
M08R15G & 0.08  &1.5 & 2.2& 0.02 & 3.60 & 69.7 $\pm$ 1.5 & 37.1 $\pm$ 1.5 & 60.2 $\pm$  1.8 \\
M08R20G & 0.08  &2.0 & 2.2& 0.05 & 3.30 & 72.1 $\pm$ 7.6 & 35.8 $\pm$ 2.6 & 57.1 $\pm$ 15.0 \\
M08R25G & 0.08  &2.5 & 2.2& 0.07 & 3.12 & 70.7 $\pm$ 3.8 & 47.2 $\pm$ 10.1& 39.9 $\pm$ 20.8 \\
M08R30G & 0.08  &3.0 & 2.2& 0.09 & 3.04 & 69.5 $\pm$ 2.1 & 52.6 $\pm$ 2.7 & 25.5 $\pm$  5.4 \\
M08R35G & 0.08  &3.5 & 2.2& 0.12 & 2.98 & 73.6 $\pm$ 4.5 & 61.4 $\pm$ 6.7 & 28.9 $\pm$  5.3 \\
\hline
M15R15G & 0.15  &1.5 & 2.1& 0.04 & 3.58 & 74.2 $\pm$ 4.0 & 39.1 $\pm$ 1.5 & 74.7 $\pm$ 13.9 \\
M15R20G & 0.15  &2.0 & 2.1& 0.08 & 3.28 & 71.0 $\pm$ 4.4 & 50.3 $\pm$ 4.7 & 32.8 $\pm$ 13.2 \\
M15R25G & 0.15  &2.5 & 2.2& 0.13 & 3.12 & 75.0 $\pm$ 3.5 & 65.6 $\pm$ 2.3 & 16.1 $\pm$  4.4 \\
M15R30G & 0.15  &3.0 & 2.2& 0.17 & 3.04 & 72.9 $\pm$ 0.9 & 69.0 $\pm$ 3.0 &  6.3 $\pm$  4.1 \\
M15R35G & 0.15  &3.5 & 2.2& 0.21 & 2.98 & 74.7 $\pm$ 1.6 & 69.8 $\pm$ 3.5 &  8.0 $\pm$  6.8 \\
\hline
M30R15G & 0.30  &1.5 & 1.8& 0.08 & 3.56 & 74.5 $\pm$ 4.2 & 42.7 $\pm$ 3.6 & 65.7 $\pm$ 19.8 \\
M30R20G & 0.30  &2.0 & 1.8& 0.16 & 3.26 & 75.1 $\pm$ 5.9 & 60.6 $\pm$ 4.9 & 25.2 $\pm$  11.1 \\
M30R25G & 0.30  &2.5 & 2.1& 0.23 & 3.10 & 78.5 $\pm$ 2.6 & 69.7 $\pm$ 2.8 & 18.2 $\pm$  10.6 \\
M30R30G & 0.30  &3.0 & 2.1& 0.31 & 3.02 & 78.2 $\pm$ 3.2 & 75.2 $\pm$ 5.0 &  6.4 $\pm$  6.0 \\
M30R35G & 0.30  &3.5 & 2.1& 0.39 & 2.90 & 79.6 $\pm$ 0.1 & 79.7 $\pm$ 0.5 &  2.7 $\pm$  1.2 \\
\hline
M60R15G & 0.60  &1.5 & 1.4& 0.14 & 3.42 & 80.2 $\pm$ 3.1 & 57.1 $\pm$  5.7& 64.2 $\pm$ 24.0 \\
M60R20G & 0.60  &2.0 & 1.3& 0.28 & 3.18 & 79.1 $\pm$ 4.1 & 69.3 $\pm$  8.9& 22.8 $\pm$ 16.2 \\
M60R25G & 0.60  &2.5 & 1.3& 0.41 & 3.02 & 76.6 $\pm$ 3.2 & 72.7 $\pm$  1.5& 9.2  $\pm$  9.8 \\
M60R30G & 0.60  &3.0 & 1.9& 0.54 & 2.92 &      $\cdots$  &      $\cdots$  &      $\cdots$   \\
M60R35G & 0.60  &3.5 & 1.9& 0.67 & 2.88 &      $\cdots$  &
$\cdots$  &      $\cdots$
\enddata
\tablecomments{See Table \ref{tbl:model_no} for the definitions of
the various symbols.}
\end{deluxetable*}

For models in which self-gravity is included, we employ the method
used by \citet{she08} to calculate the gravitational potential from
the perturbed gas surface density, $\Sigma-\Sigma_0$.\footnote{In
our self-gravitating models, gravity of the initial gaseous disk
with surface density $\Sigma_0$ is neglected in order to make the
initial rotation curve identical to that in the non-self-gravitating
counterparts.} This method is based on \citet{kal71}'s scheme that
is efficient on a logarithmically-spaced radial grid. As a softening
parameter, we take $H/r=0.1$ that allows for non-zero thickness of
the disk in the potential calculation.  In our models, the gaseous
disk initially has a Toomre stability parameter of
\begin{equation}\label{eq:QT}
Q_T \equiv \frac{\kappa\cs}{\pi G\Sigma} \approx
2.1\left(\frac{r}{10 \kpc}\right)^{-1}
\left(\frac{\Sigma}{10\Surf}\right)^{-1},
\end{equation}
so that it is gravitationally stable.  However, nuclear rings that
form at around $r\sim1\kpc$ would achieve sufficient density to
be unstable when the bar is strong.

We solve the basic equations of ideal hydrodynamics using the CMHOG
code in a frame corotating with the bar (Paper I).  CMHOG is
third-order accurate in space and has very little numerical
diffusion \citep{pin95}. We resolve the central regions with high
accuracy by setting up a logarithmically-spaced cylindrical grid
extending from 0.04 kpc at the inner boundary to 16 kpc at the outer
boundary. The size of the inner boundary taken in the present models
is twice larger than that in Paper I: we checked that this makes
negligible differences in the properties of bar substructures except
for the mass inflow rate across the inner boundary that becomes
larger, by $\sim10\%$ on average, in the current models. The number
of zones in our models is 1024 in the radial direction and 535 in
the azimuthal direction that covers the half-plane. The
corresponding spatial resolution is $\Delta r=0.23$, 5.86, and 93.8
pc at the inner boundary, $r=1\kpc$, and the outer boundary,
respectively. We adopt the outflow and continuous boundary
conditions at the inner and outer boundaries, respectively, while
taking the periodic conditions at the azimuthal boundaries. The bar
potential is slowly turned on over one bar revolution time
$2\pi/\Omb=186$ Myr to minimize transients in the flows caused by
its sudden introduction. All the models are run until 1 Gyr.


\begin{figure}
\hspace{0.5cm}\includegraphics[width=8cm]{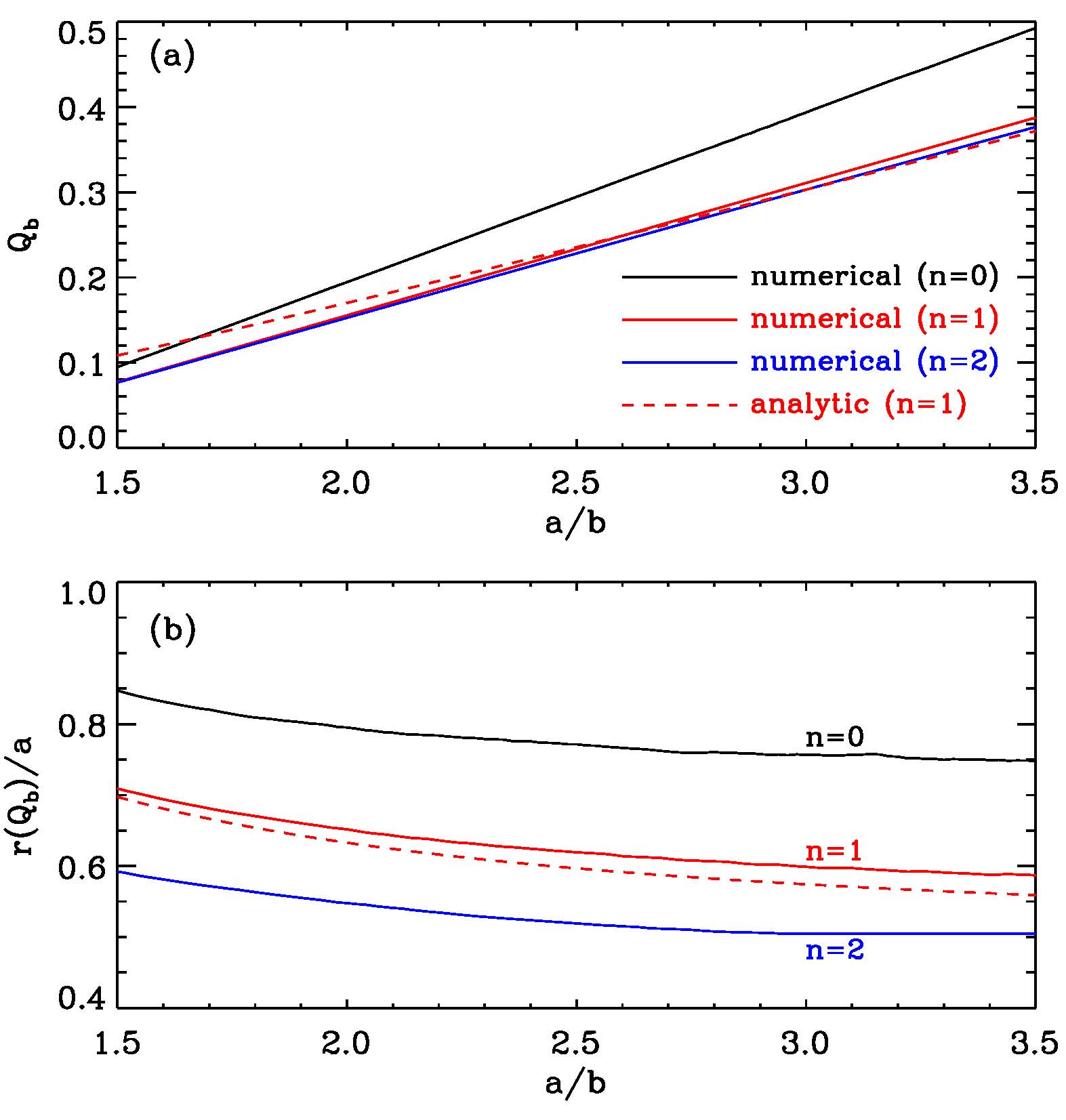}
\caption{Dependence of the bar strength $\Qb$ and the maximum-torque
radius $\RQ$ on the bar aspect ratio $\R=a/b$ for $\fbar=0.3$. The
solid curves plot the numerical results for our galaxy models with
$n=0, 1, 2$ Ferrers bars, while the dashed lines are for the
analytic results assuming flat rotation. \label{fig:Qb_rmax}}
\end{figure}

\section{Bar Strength\label{sec:Qb}}

One of the key parameters that govern the gas dynamics in barred
galaxies is the bar strength which depends on both $\fbar$ and $\R$.
In this section, we evaluate the bar strength of our numerical
models that will be used to analyze the properties of bar
substructures in Section 4.  We also provide fitting formulae for
the bar strength and the position of the maximum bar torque for
future purposes.

It has often been customary to measure the bar strength using the
dimensionless parameter $\Qb$ defined by
\begin{equation}\label{eq:Qb}
\Qb \equiv \left.\frac{F_T(r,\phi)}{F_R(r)}\right|_{\rm max},
\end{equation}
where $F_T=-(\partial\Pbar/\partial \phi)/r$ is the tangential force
due to the non-axisymmetric bar potential and $F_R = -(1/2\pi)
\int_0^{2\pi} \partial \Ptot/\partial r d\phi$ is the
azimuthally-averaged radial force (e.g., \citealt{com81,lau02,
blo04,lau04,lau06,pee06,com09,com10}).\footnote{The non-axisymmetric
torque is sometimes represented by $Q_g$ that includes a
contribution from spiral arms (e.g., \citealt{dur09,com10}).  In our
models, $\Qb=Q_g$ since no spiral arm is considered.} Physically,
$\Qb$ corresponds to the maximum bar torque applied to a gaseous
material in orbital motion relative to its specific kinetic energy.
\citet{blo04} found that $\Qb$ is deeply related to the class of
barred-spiral galaxies such that SA galaxies have $\Qb\simlt 0.1$,
while $\Qb\simgt 0.15$ for SB galaxies.

Since the bar torque is stronger for a more massive and elongated
bar, $\Qb$ should be an increasing function of $\fbar$ and $\R$. It
is straightforward to calculate $\Qb$ values of our galaxy models as
well as the radius $\RQ$ where $|F_T|/F_R$ is maximized. Figure
\ref{fig:Qb_rmax} plots as solid lines the resulting $\Qb$ and $\RQ$
as functions of $\R$ for the Ferrers bar with $\fbar=0.3$ and $n=0,
1, 2$. Note that when $\fbar=0.3$, $\Qb$ for $n=2$ is not much
different from the case with $n=1$, while $\RQ$ becomes smaller by
$\sim16\%$ as $n$ increases from 1 to 2. Also plotted in Figure
\ref{fig:Qb_rmax} as dashed lines are the analytic expressions
derived in Appendix for the Ferrers bars with $n=1$ assuming flat
rotation, which agrees with the numerical results within less than
15\%.  The difference between the numerical and analytic results for
$n=1$ is of course due to the fact that the rotation curves in our
models are not strictly flat but decrease slowly with $r$ in the
regions of the maximum bar torque, tending to increase $\RQ$
compared to the case with exactly flat rotation. Columns (5) and (6)
of Tables \ref{tbl:model_no} and \ref{tbl:model_sg} list $\Qb$ and
$\RQ$ of our numerical models.

By varying $\fbar$, $\R$, and $n$, we find that
\begin{equation}\label{eq:Qbnum}
\Qb = \left\{\begin{array} {l l@{\ }c@{\,}l}
      0.58\fbar^{0.89}(a/b-1),  & \textrm{for} & n=0, \\
      0.44\fbar^{0.87}(a/b-1),  & \textrm{for} & n=1, \\
      0.38\fbar^{0.79}(a/b-1),  & \textrm{for} & n=2,
      \end{array}\right.
\end{equation}
and
\begin{equation}\label{eq:Rmaxnum}
\frac{\RQ}{a} = \left\{\begin{array} {l l@{\ }c@{\,}l}
      1.024 - 0.161(a/b) + 0.024(a/b)^2,  & \textrm{for} & n=0, \\
      0.934 - 0.196(a/b) + 0.028(a/b)^2,  & \textrm{for} & n=1, \\
      0.817 - 0.197(a/b) + 0.031(a/b)^2,  & \textrm{for} & n=2,
      \end{array}\right.
\end{equation}
are good fits, within 5\%, to the numerical results. While $\Qb$ is
linearly proportional to $\R$, $\RQ$ depends very weakly on $\R$ and
is independent of $\fbar$. The reason for $\Qb$ varying less steeply
than $\fbar^{1.0}$ is that large $\fbar$ increases the rotational
velocity slightly near the regions where the bar torque achieves its
maximum.

\section{Simulation Results\label{sec:result}}

Paper I presented gas dynamical evolution of Model M30R25N with
$\Qb=0.23$ ($\fbar=0.3$ and $\R=2.5$) in detail. Evolution of other
models with differing $\Qb$ are qualitatively similar except that
dust lanes and/or nuclear rings are absent if $\fbar$ and/or $\R$
are too large.  In this section, we first briefly describe the main
evolutionary features and the conditions for the existence of dust
lanes and nuclear rings, and then present detailed analyses of the
properties of bar substructures formed in our models.

\subsection{Overall Morphology\label{sec:evol}}

\begin{figure*}
\hspace{0.5cm}\includegraphics[angle=0, width=17cm]{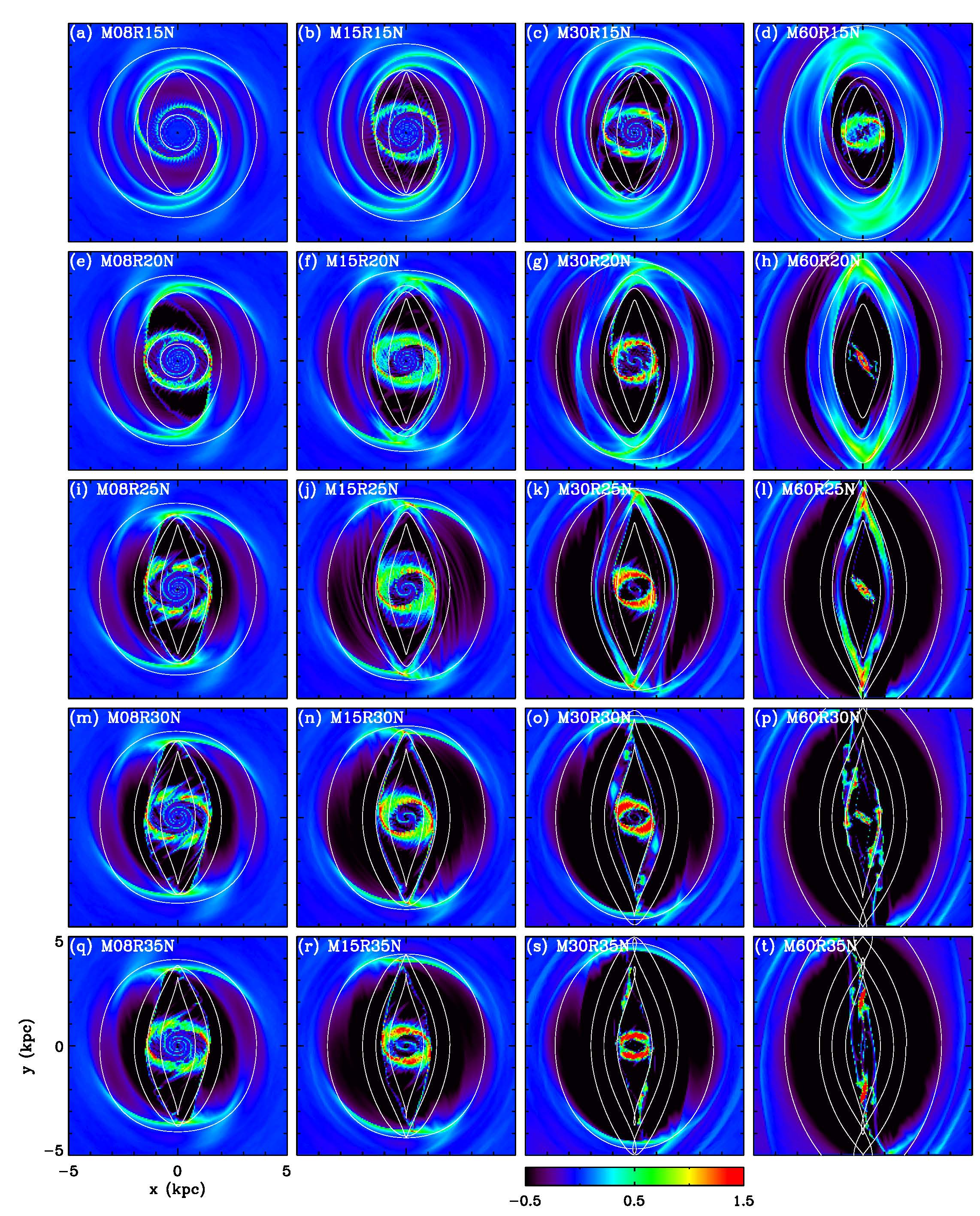}
\caption{Logarithm of the gas surface density at $t = 0.3\Gyr$ for
non-self-gravitating models. Each panel shows the inner $\pm5\kpc$
regions where the bar is oriented vertically along the $y$-axis.
Each row corresponds to models with $\R=1.5$, 2.0, 2.5, 3.0, and 3.5
from top to bottom, while each column is for models with
$\fbar=0.08$, 0.15, 0.3, and 0.6 from left to right. Solid lines in
each panel draw $\xone$-orbits that cut the $x$-axis at $x_c=0.8$,
1.4, 2.0, and $3.6\kpc$. Color bar labels $\log(\Sigma/\Sigma_0)$.
\label{fig:all_no}}
\end{figure*}

\begin{figure*}
\hspace{0.5cm}\includegraphics[angle=0, width=17cm]{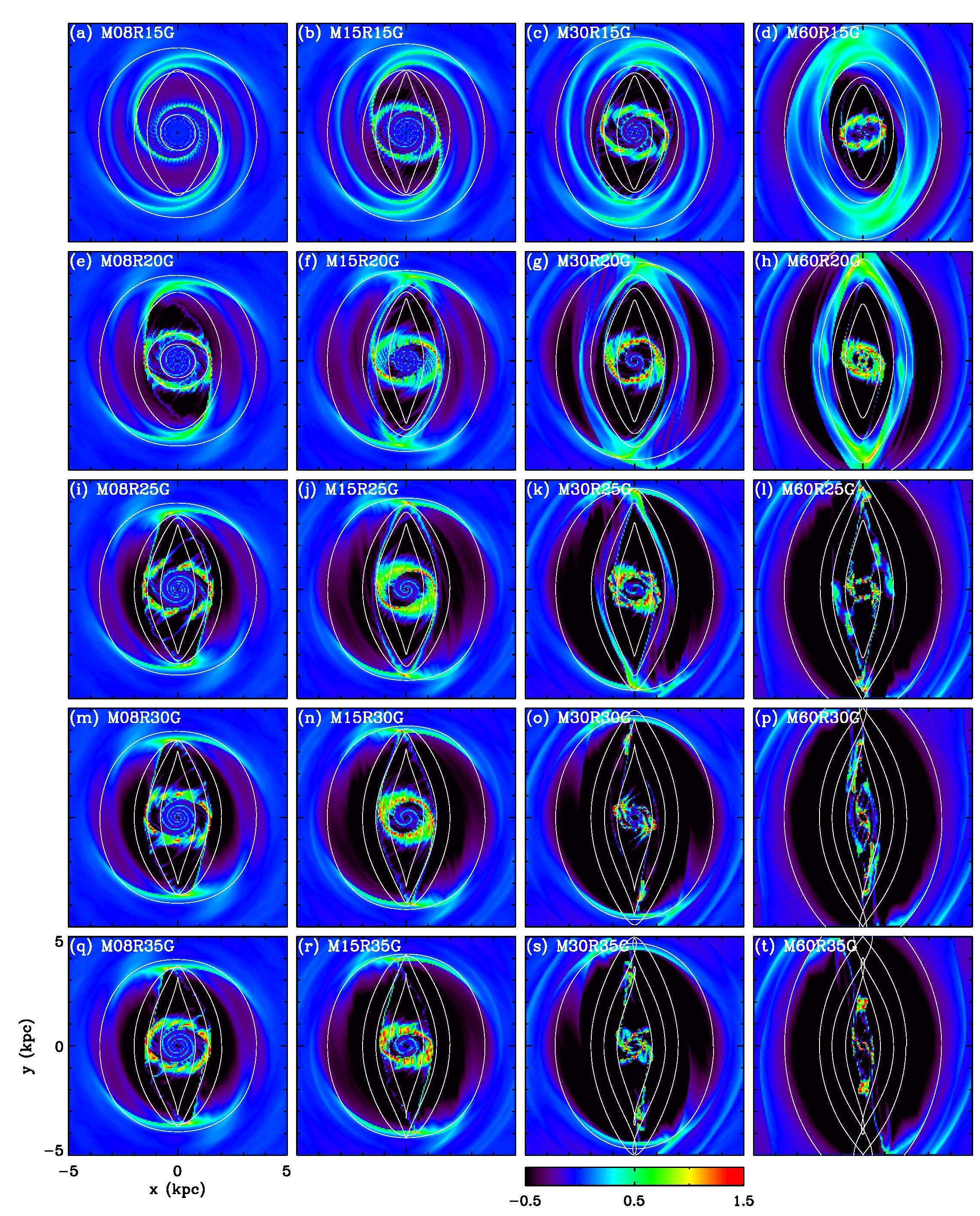}
\caption{Same as Figure \ref{fig:all_no}, but for self-gravitating
models. \label{fig:all_sg}}
\end{figure*}

\begin{figure*}
\hspace{0.5cm}\includegraphics[angle=0, width=17cm]{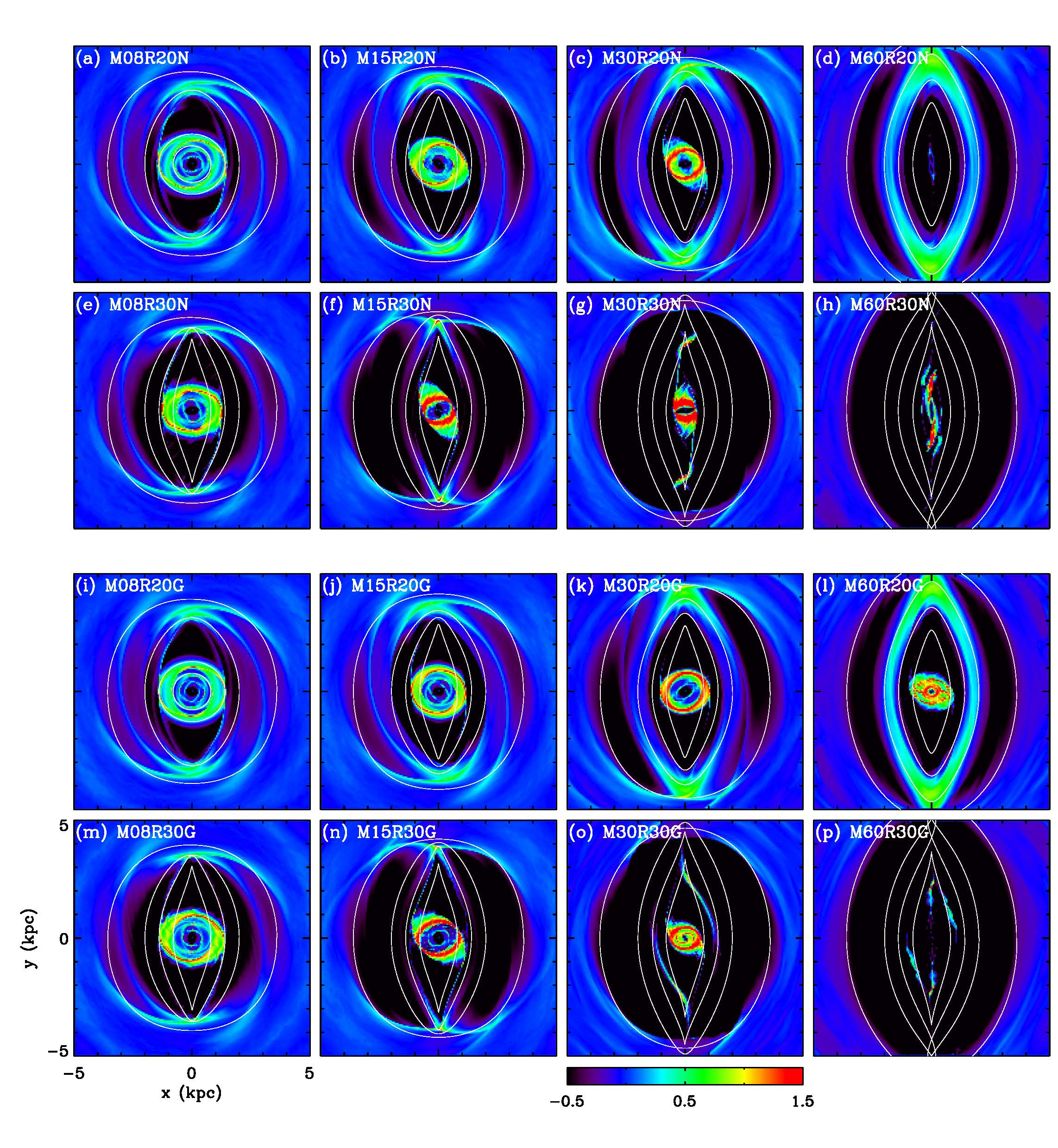}
\caption{Logarithm of the gas surface density at $t = 0.8\Gyr$ for
all models with $\R=2.0$ and 3.0. The two upper rows plot the
non-self-gravitating models, while the two bottom rows are for the
self-gravitating models. Solid lines in each panel draw
$\xone$-orbits that cut the $x$-axis at $x_c=0.8$, 1.4, 2.0, and
$3.6\kpc$. Color bar labels $\log(\Sigma/\Sigma_0)$.
\label{fig:t800}}
\end{figure*}

As the non-axisymmetric bar potential is slowly turned on,
initially-circular gas orbits are perturbed due to the bar torque,
creating overdense ridges at the downstream side of the bar major
axis. As the amplitude of the bar potential grows further, the
ridges soon develop into dust-lane shocks.  Dust-lane shocks are
found unstable to the wiggle instability identified by \citet{wad04}
(see also \citealt{kim06}) and form clumps with high vorticty (see
Paper I for more detailed description).  Gas loses angular momentum
at the shocks, flows radially inward, and forms a nuclear ring at
the position where the centrifugal force balances the external
gravity. Sonic perturbations launched from the ring propagate inward
and excite $m=2$ nuclear spirals in the nuclear regions. With
$\MBH=7\times10^7\Msun$, nuclear spirals that persist in our models
are all trailing.

Figures \ref{fig:all_no} and \ref{fig:all_sg} display snapshots of
gas surface density in logarithmic scale at $t=0.3\Gyr$ from all
non-self-gravitating and self-gravitating models, respectively.
Figure \ref{fig:t800} plots density distributions at $t=0.8\Gyr$ for
both non-self-gravitating (two upper rows) and self-gravitating (two
bottom rows) models with $\R=2.0$ and 3.0.  The inner $\pm5 \kpc$
regions are shown. The bar is oriented vertically along the
$y$-axis. The solid lines in each panel plot $\xone$-orbits that cut
the $x$-axis at $x_c=0.8, 1.4, 2.0$, and $3.6\kpc$ under the total
external potential. It is apparent that gas responses inside the
outermost $\xone$-orbit with $x_c=3.6\kpc$ are quite dramatic, while
the outer regions are relatively unperturbed (e.g., \citealt{ks12}).

At $t=0.3\Gyr$, all the models with $\Qb<0.5$ ($\fbar\leq 0.3$ or
$\R\leq2.5$) contain well-defined dust lanes that follow an
$\xone$-orbit reasonably well, although they become weaker at
$t=0.8\Gyr$. On the other hand, Models with $\Qb > 0.5$ do not
possess features that resemble observed dust lanes. In these
high-$\Qb$ models, the bar torque is so strong that the gas inside
the outermost $\xone$-orbit loses most of its initial angular
momentum even in the developing stage of dust lanes ($t\sim
0.13\Gyr$ for Model M60R35N). With the shocked gas rapidly lost to
the nuclear regions and subsequently to the inner boundary, there
remains no material available to support the dust lanes in the bar
regions. In these models, the bar regions are occupied by gas blobs
in a filamentary shape that are repeatedly stretched and folded on
the course of orbital motions along $\xone$-orbits about the center.
Comparison between Figures \ref{fig:all_no} and \ref{fig:all_sg}
shows that the shape of dust lanes is not much affected by
self-gravity.

\begin{figure}
\hspace{0.2cm}\includegraphics[angle=0, width=8.5cm]{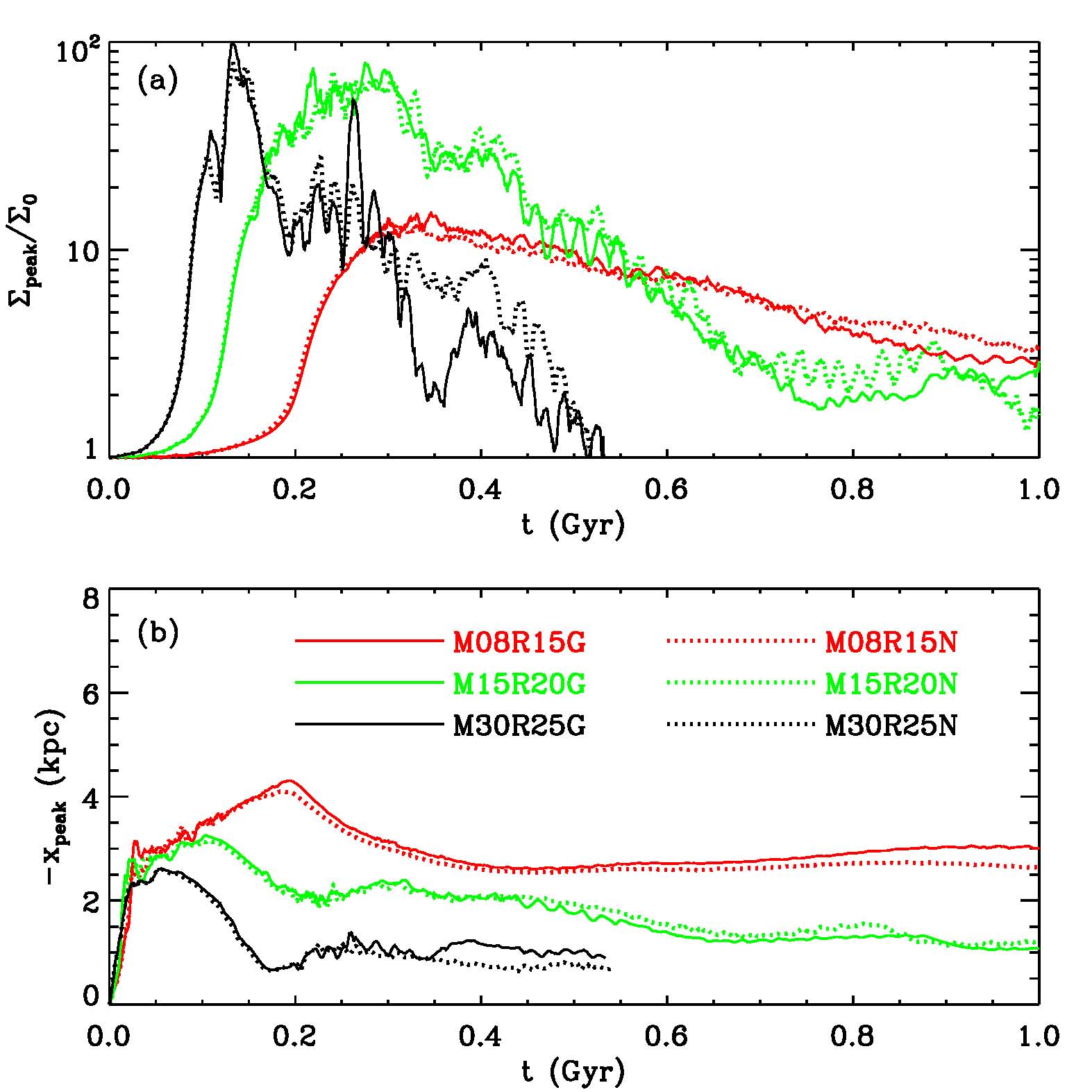}
\caption{Temporal evolution of (a) the peak gas surface density
$\pSig$ and (b) the $x$-coordinate $\xp$ of the dust lanes at
$r=1.5\kpc$ for Models M08R15, M15R20, and M30R25. The solid and
dotted lines are for the self-gravitating and non-self-gravitating
models, respectively. For Models M30R25N and M30R25G, only the
results up to $t=0.54\Gyr$ are shown, after which dust lanes are too
weak to be defined well. \label{fig:dust}}
\end{figure}

Figures \ref{fig:all_no} also shows that conventional $\xtwo$-type
rings, that is, rings whose shape is similar to $\xtwo$-orbits,
exist when $\fbar\leq0.3$ or $\R\leq1.5$, in non-self-gravitating
models. Rings are quite clumpy even in non-self-gravitating models
due to addition of clumps produced by the wiggle instability at the
dust-lane shocks. Nuclear rings in Models M60R20N and M60R25N are
highly inclined relative to the $x$-axis at $t=0.3\Gyr$ and
gradually precess to align their long axes to the bar major axis,
eventually becoming $\xone$-type rings, although they become very
weak at late time (e.g., Fig.\ \ref{fig:t800}d). In Models M60R30N
and M60R35N, orbits of gas in the nuclear regions are highly
transient to be considered as rings.  In self-gravitating models, on
the other hand, rings that form are all of the $\xtwo$ type (Fig.\
\ref{fig:all_sg}). In Models M60R30G and M60R35G, rings at
$t=0.3\Gyr$ are so small that they become weaker with time by losing
mass directly to the central hole, disappearing at $t\sim0.75\Gyr$.
For models that form an $\xtwo$-type ring, rings are larger in size
in self-gravitating models than in non-self-gravitating models
(Fig.\ \ref{fig:t800}). We will explain how gas orbits are affected
by the bar potential and self-gravity in Section \ref{sec:ring}
below.

As $\Qb$ decreases, angular momentum loss at the shocks becomes
smaller, resulting in a larger nuclear ring. Dust lanes are
correspondingly located further downstream from the bar major axis
since their inner ends are always attached to a nuclear ring. While
dust lanes become weaker with time after the peak strength is
attained, nuclear rings remain strong until the end of the runs
(e.g., Fig.\ \ref{fig:t800}), suggesting that they are long-lived
features (e.g., \citealt{all06,sar07,com10}).  In Models M08R15N and
M08R15G with $\Qb=0.02$, the bar torque is so weak that the amount
of gas moving in along the dust lanes is not sufficient to form an
appreciable ring until the end of the runs.

The presence of nuclear spirals is deeply related to the shape and
size of a nuclear ring.  There is no chance to possess nuclear
spirals in models with an eccentric $\xone$-type ring.  Even in
models with an $\xtwo$-type ring, nuclear spirals are absent if a
nuclear ring is too small (as in Models M30R35N, M30R35G, M60R20G,
and M60R25G) to provide enough space for coherent structures to
grow. In Models M08R15N and M08R15G, on the other hand, the
inflowing gas excites trailing spiral waves at $r\sim 1\kpc$ that
propagate radially inward and turn into nuclear spirals. We defer
the more detailed discussion on temporal evolution of nuclear
spirals to Section \ref{sec:nsp}.

\subsection{Dust Lanes}\label{sec:dust}

To quantify the strength and displacement of dust lanes from the bar
major axis, we confine to the regions with $r=1.5\kpc$ and measure
the $x$-coordinate, $\xp$, of the position where gas surface density
is maximized at each time. Figure \ref{fig:dust} plots the temporal
changes of the peak surface density $\pSig$ as well as $\xp$ at
$r=1.5\kpc$ in Models M08R15, M15R20, and M30R25.  The solid and
dotted lines are for the self-gravitating and non-self-gravitating
models, respectively. In Model M30R25, $\pSig$ rises with time and
peaks at $t\sim0.13-0.15\Gyr$ after which it decreases with time as
the bar regions become increasingly evacuated due to gas infalls to
the ring.  At early time when the bar potential is weak, dense
ridges form at $\xp\sim-2.5\kpc$ which move closer to the bar major
axis as they turn to dust-lane shocks and stay at $\xp\sim-1 \kpc$.
In these models with $\Qb=0.25$, the decay of the dust lanes is
relatively rapid that they become almost invisible after
$t=0.54\Gyr$. The temporal behaviors of $\pSig$ and $\xp$ are
qualitatively the same for other models with different bar strength,
although the dust lanes tend to decay more slowly and locate farther
from the bar major axis as $\Qb$ decreases. Note that while the
strength of dust lanes varies with time considerably, their location
and shapes do not change much after $t=0.25\Gyr$. Note also that
self-gravity does not make much changes in $\pSig$ and $\xp$. This
is because dust lanes with typical density $\pSig\sim100\Surf$ at
$r=1.5\kpc$ have $Q_T\sim 1.4$, larger than unity (e.g., Eq.\
[\ref{eq:QT}]). In addition, dust lanes are known to possess strong
velocity shear, about 10 times larger than the velocity shear in
disks at large (Paper I), which tends to reduce the mass-collecting
effect of self-gravity in dust lanes.

\begin{figure*}
\hspace{0.5cm}\includegraphics[angle=0, width=17cm]{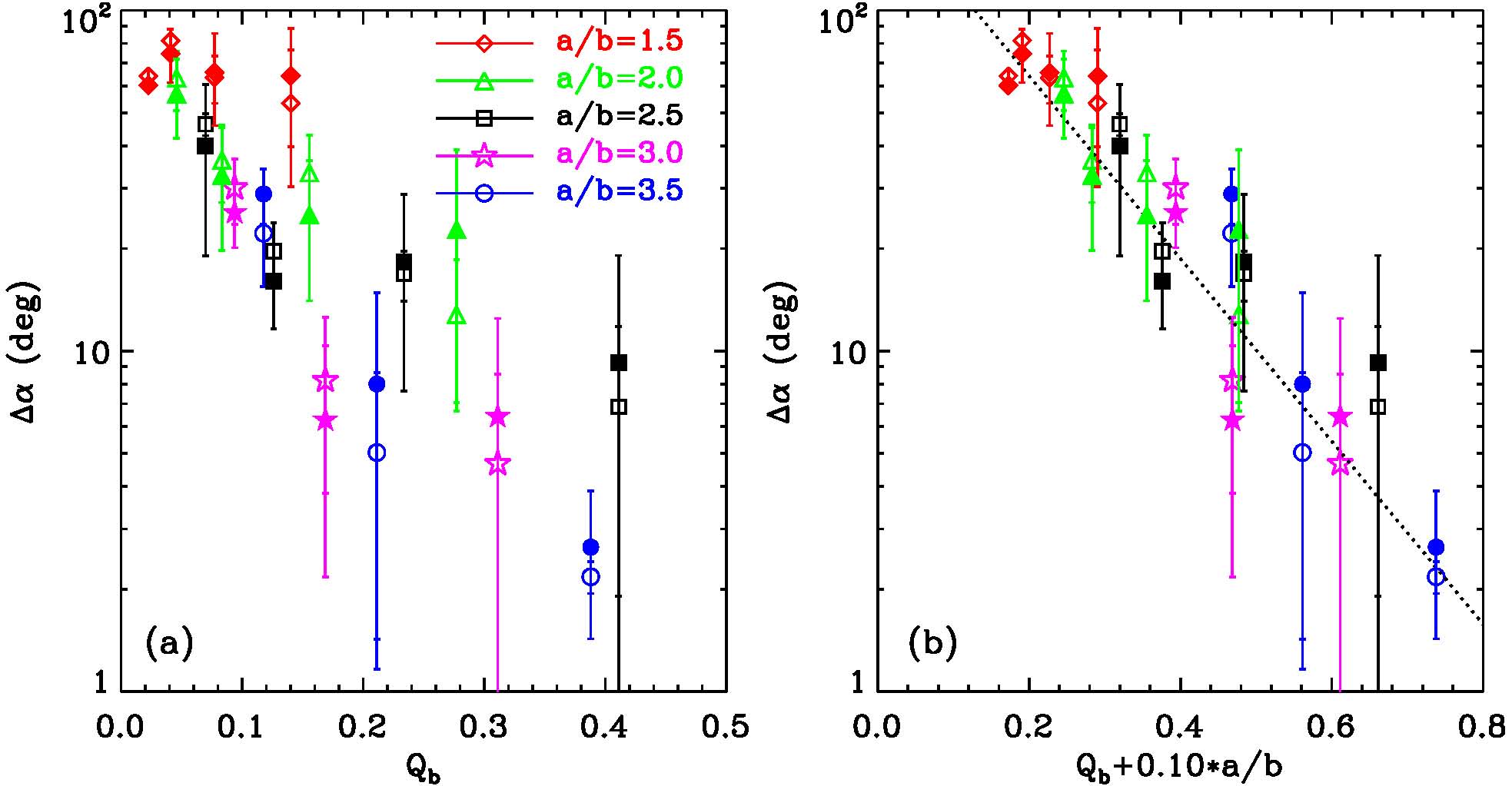}
\caption{Curvatures $\Dalp$ of dust lanes as functions of (a) $\Qb$
alone and (b) a linear combination of $\Qb$ and $\R=a/b$ for all
models that posses dust lanes.  Various symbols give the mean values
of $\Dalp$ averaged over $t=0.25$--$0.35\Gyr$, while the errorbars
indicate the standard deviations.  In both panels, filled and open
symbols are for self-gravitating and non-self-gravitating models,
respectively. The dotted line in (b) draws the best fit, expressed
in equation (\ref{eq:dalpfit}). \label{fig:dalp}}
\end{figure*}

\begin{figure*}
\hspace{0.5cm}\includegraphics[angle=0, width=17cm]{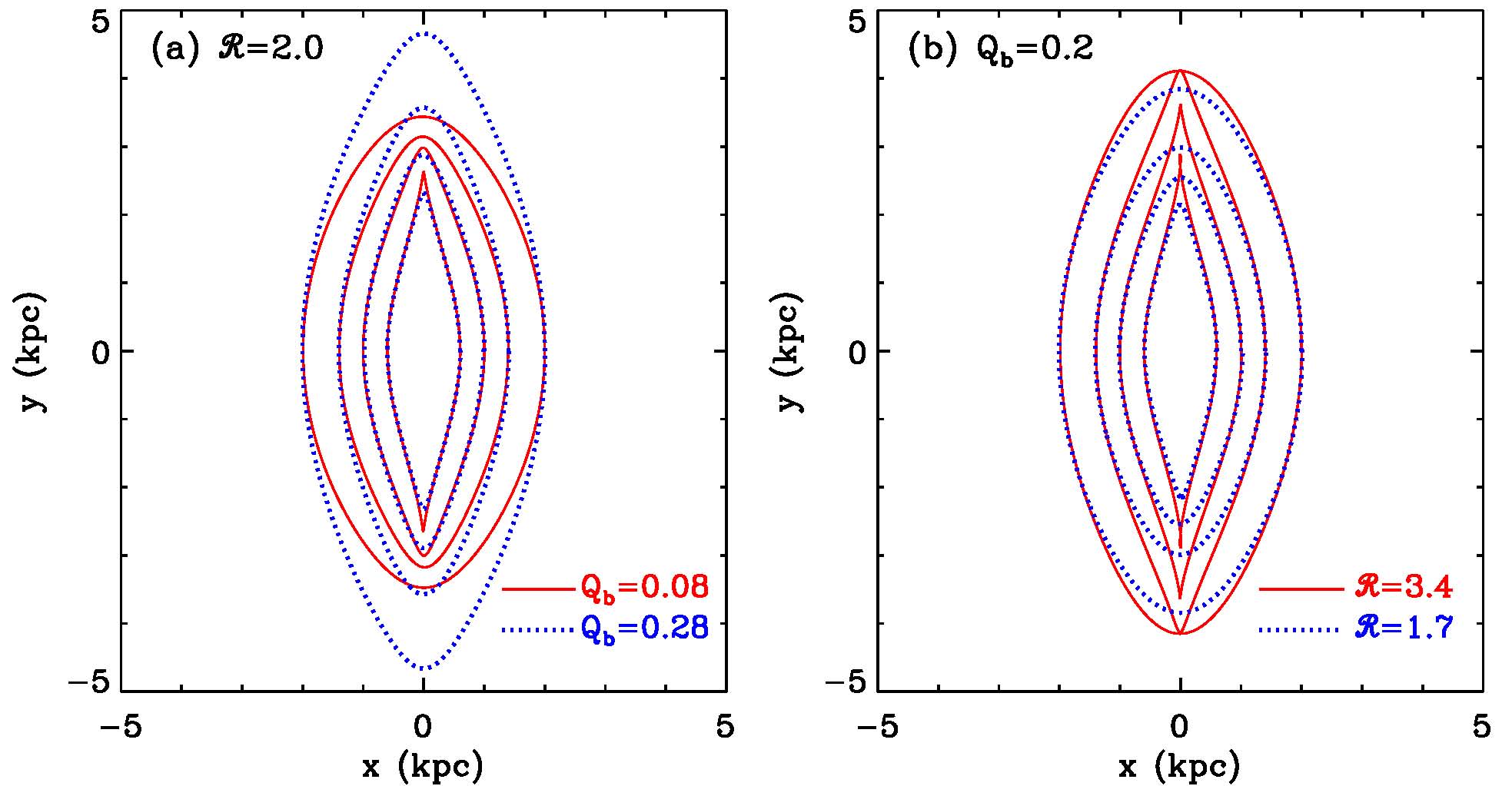}
\caption{Dependence of $\xone$-orbits with $x_c=0.6, 1.0, 1.4,
2.0\kpc$ (a) on $\Qb$ when $\R=2.0$ is fixed, and (b) on $\R$ when
$\Qb=0.2$ is fixed.  Note that the parts $\xone$-orbits in a
quadrant are more curved as $x_c$ increases or $\R$ decreases.
\label{fig:orb_comp}}
\end{figure*}

\citet{com09} defined the dimensionless curvature of a dust-lane
segment as
\begin{equation}\label{eq:dalp}
\Dalp = \frac{\alpha_1 - \alpha_2}{|\mathbf{r}_1 - \mathbf{r}_2|}
\RQ,
\end{equation}
where $\alpha_{1,2}$ and $\mathbf{r}_{1,2}$ indicate the tangent
angle to, and the position vector of, the inner and outer ends of
the segment, respectively. This extends the definition of $\Dalp$ in
\citet{kna02} to take into account the length of the host bar. To
measure $\Dalp$, \citet{com09} selected a constant-curvature part of
a dust lane in each galaxy, with the determination of both ends
relying on visual inspection. In order to obtain $\Dalp$
unambiguously from our numerical models, we instead confine to a
dust-lane segment bounded by the fixed position angles
$\phi_{1,2}=135^\circ, 100^\circ$ (or, $\phi_{1,2}=-45^\circ,
-80^\circ$) from the positive $x$-axis, and measure $\alpha_{1,2}$
and $\mathbf{r}_{1,2}$ at both ends. Columns (7)--(9) of Tables
\ref{tbl:model_no} and \ref{tbl:model_sg} list the mean values
(together with standard deviations) of $\alpha_1$, $\alpha_2$, and
$\Dalp$ averaged over $t=0.25$--$0.35\Gyr$ for models that possess
dust lanes.  While $\alpha_2$ varies rather sensitively with $\Qb$,
$\alpha_1\sim70^\circ$--$80^\circ$ does not change much, indicating
that $\Dalp$ is determined primarily by the tangent angle at the
outer end of dust lanes.

\begin{figure*}
\hspace{0.5cm}\includegraphics[angle=0, width=17cm]{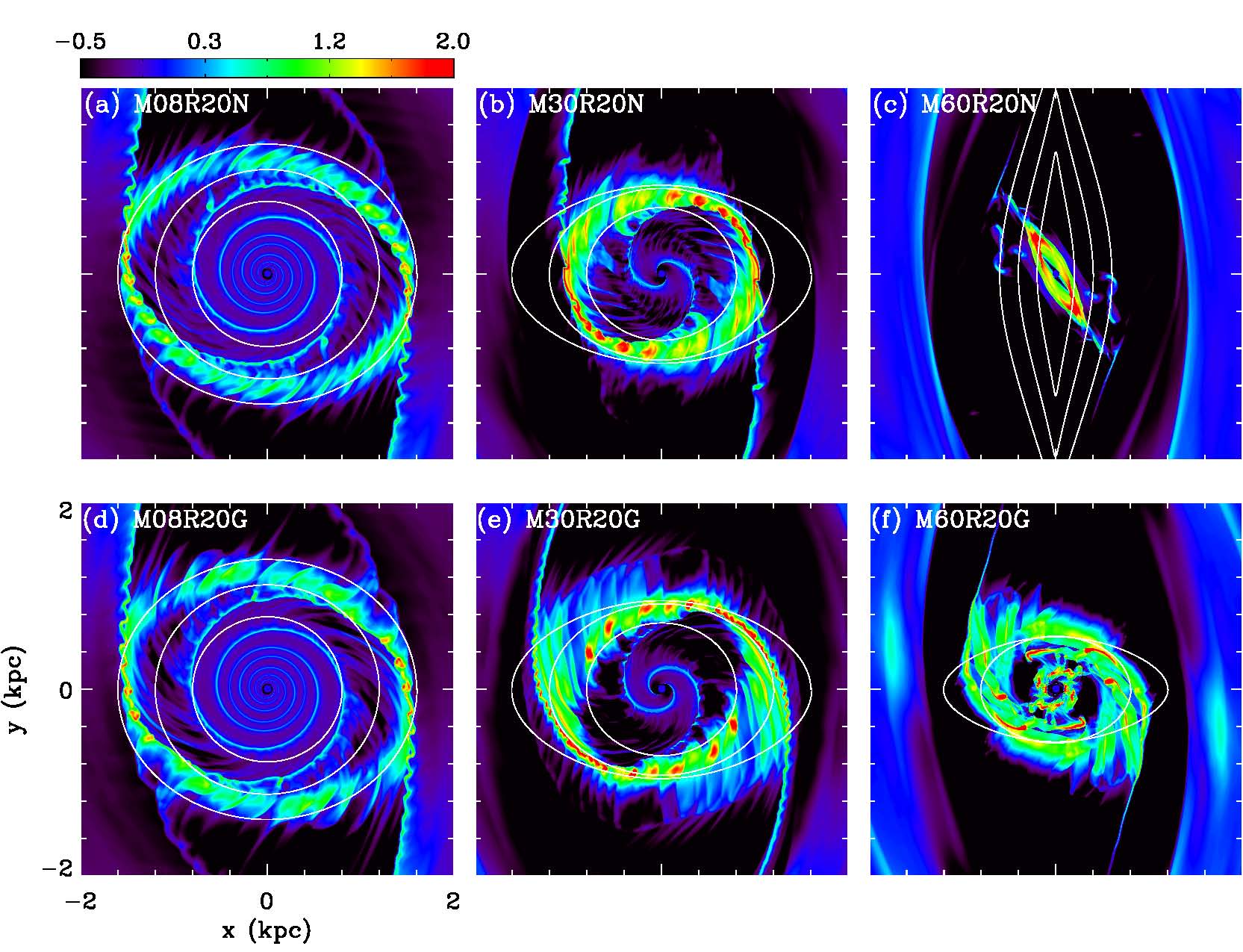}
\caption{Snapshots of logarithm of surface density at $t=0.3\Gyr$ in
the inner $2\kpc$ regions of non-self-gravitating models (a)
M08R20N, (b) M30R20N, and (c) M60R20N, and self-gravitating models
(d) M08R20G, (e) M30R20G, and (f) M60R20G. The solid lines draw
$\xtwo$-orbits that cut the $x$-axis at $x_c=0.8, 1.2, 1.6\kpc$ in
(a), (b), (d), and (e), $\xone$-orbits with $x_c=0.2, 0.4, 0.6\kpc$
in (c), and $\xtwo$-orbits with $x_c=0.8, 1.2\kpc$ in (f). Color bar
labels $\log(\Sigma/\Sigma_0)$. \label{fig:inner}}
\end{figure*}

Figure \ref{fig:dalp}a plots $\Dalp$ measured from our numerical
models as a function of $\Qb$.  Filled and open symbols are for
self-gravitating and non-self-gravitating models, respectively.
These agree within error bars, indicating that self-gravity does not
affect the shape of dust lanes much. In general, $\Dalp$ decreases
with increasing $\Qb$, consistent with the prediction of
\citet{ath92b} (see also \citealt{kna02}). Note however that there
is a considerable scatter in $\Dalp$ for given $\Qb$. \citet{com09}
found that the spread in the data points can be much reduced by
fitting them using a linear combination of $\Dalp$ and $\R$. We
follow the same procedure and find the best fit
\begin{equation}\label{eq:dalpfit}
\Qb+0.10\R = 0.87 - 0.37 \log \Dalp,
\end{equation}
for all models, which is plotted as a dotted line in Figure
\ref{fig:dalp}b. The chi-square measure of equation
(\ref{eq:dalpfit}) is about 3 times smaller than that of the best
fit using only $\Qb$. Equation (\ref{eq:dalpfit}) implies that more
elongated bars lead to more straight dust lanes.

The behavior of $\Dalp$ upon $\Qb$ and $\R$ given in equation
(\ref{eq:dalpfit}) can be qualitatively understood from the facts
that dust lanes roughly follow $\xone$-orbits and that they move
closer to the bar major axis as $\Qb$ increases.  Figure
\ref{fig:orb_comp} illustrates the changes in the shapes of
$\xone$-orbits with $x_c=0.6, 1.0, 1.4, 2.0\kpc$ as $\Qb$ or $\R$
varies. For fixed $\R$, the parts of $\xone$-orbits, say, in the
second or fourth quadrant where dust lanes are located, are clearly
more curved as $x_c$ increases, explaining larger $\Dalp$ in models
with smaller $\Qb$. For fixed $\Qb$, $\xone$-orbits are more acute
near the $y$-axis under a more elongated bar (i.e., larger $\R$),
while they are largely independent of $\R$ near the $x$-axis. This
makes $\alpha_1$ almost unchanged with $\R$, while causing
$\alpha_2$ to increase with $\R$, resulting in smaller $\Dalp$ for
larger $\R$.

\subsection{Nuclear Ring}\label{sec:ring}

A conventional wisdom about nuclear rings is that they form near the
ILRs as a result of resonant interactions of gas with the background
potential (e.g., \citealt{com96,but96}).  In this subsection, we use
our numerical models to study what controls the ring formation and
its size when there is a single ILR. Paper I showed that the mass of
a central BH (and thus the number and locations of ILRs) does not
much affect the physical properties of nuclear rings that form, so
that the results presented below hold also for models with two ILRs.
We first describe the results of non-self-gravitating models and
then discuss the effect of self-gravity on nuclear rings.

\begin{figure}
\hspace{0.2cm}\includegraphics[angle=0, width=8.5cm]{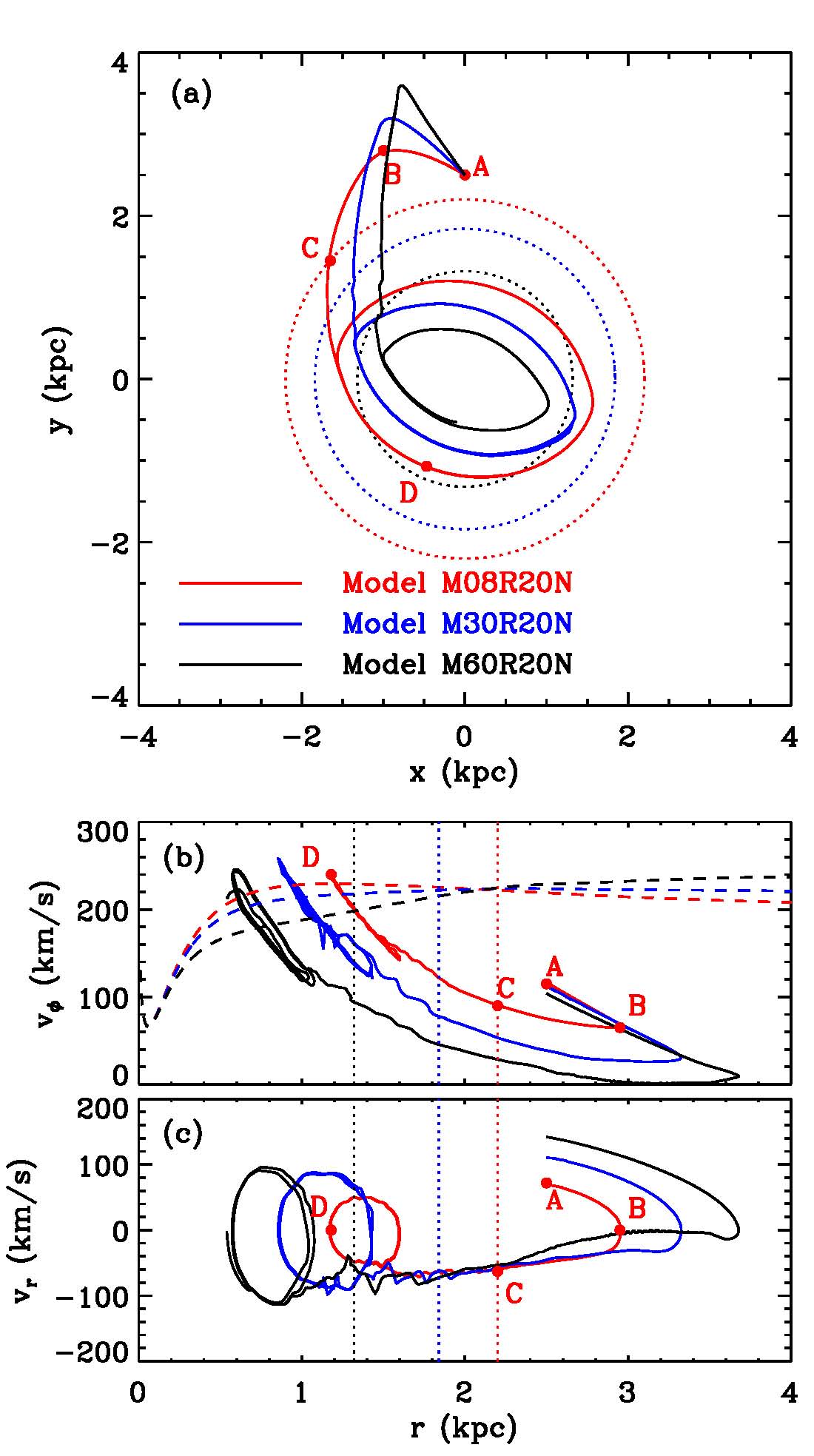}
\caption{(a) Instantaneous streamlines of the gas that starts from
Point A $(x,y)=(0, 2.5\kpc)$ in Model M08R20N at $t=0.25\Gyr$ (red),
Model M30R20N at $t=0.15\Gyr$ (blue), and Model M60R20N at
$t=0.12\Gyr$ (black) when a nuclear ring is beginning to form. The
dotted circles mark the locations of the ILR at $\ILR=2.2, 1.8, 1.3
\kpc$ for Models M08R20N, M30R20N, M60R20N, respectively. (b and c)
The variations of the azimuthal and radial velocities of the gas
along the paths shown in (a).  The ILRs  are indicated as vertical
lines. In (b), the dashed lines draw the equilibrium circular
velocity in each model. \label{fig:stline}}
\end{figure}

\subsubsection{Non-self-gravitating Models}\label{sec:ring_no}

Inspection of Figure \ref{fig:all_no} reveals that the ring shape
and its position depend on $\fbar$ and/or $\R$.  To show this more
clearly, the upper panels of Figure \ref{fig:inner} zoom in on the
central $\pm 2\kpc$ regions and plots density distributions in
logarithmic scale at $t=0.3\Gyr$ together with a few $\xone$- or
$\xtwo$-orbits for selected models with $\R=2.0$. It is apparent
that Models M08R20N and M30R20N have an $\xtwo$-type nuclear ring,
while the ring in Model M60R20N is inclined in the galaxy plane with
respect to the $x$-axis. Rings become smaller with increasing
$\fbar$, while Model M60R20N has a very eccentric ring.

To illustrate how nuclear rings form in our models, we go back to
early time when they were beginning to shape. Figure
\ref{fig:stline} plots the instantaneous streamlines of the gas that
starts from Point A marked at $(x, y) = (0,2.5\kpc)$ as well as the
variations of the rotational and azimuthal velocities in the
inertial frame along the streamlines for Model M08R20N at
$t=0.25\Gyr$ (red), Model M30R20N at $t=0.15\Gyr$ (blue), and Model
M60R20N at $t=0.12\Gyr$ (black). In each panel, the dotted circles
or vertical lines mark the ILR located at $\ILR=2.2, 1.8, 1.3\kpc$
for Models M08R20N, M30R20N, M60R20N, respectively. Dashed lines in
Figure \ref{fig:stline}b draw the equilibrium rotation curve,
averaged between on the major and minor axes of the bar, in each
model. A similar plot is given in Figure 5 of Paper I for models
with no BH and in Figure 5 of \citet{ks12} for magnetized models.

Outside the ILR, the bar provides a negative torque for the gas,
producing dust-lane shocks.  In Model M08R20N, the gas that passes
through Point A hits the shocks at Point B. By losing angular
momentum there, it starts to move radially in and crosses the ILR at
Point C.  The radial velocity of the inflowing gas at Point C is
rather large at $\sim63\kms$. The associated ram pressure is larger,
by about a factor of 40 and 4, respectively, than the thermal
pressure and the bar torque at the ILR that may try to stop the
inflowing motion of the gas. Thus, the inflowing gas is not halted
at the ILR and continues to move inward.  At the same time, the gas
rotates gradually faster at the expense of the gravitational
potential energy.  It is at Point D that the gas achieves its
rotational velocity comparable to the equilibrium value and
subsequently makes a closed-loop orbit, finally forming a nuclear
ring. The mean radius of the ring in Model M08R20N is
$\Rring\sim1.4\kpc$, well inside the ILR. Models with a more massive
bar forms a ring closer to the center, owing to a larger loss of
angular momentum.  All rings form at the position where the
centrifugal force balances the external gravity. These imply that
the formation of nuclear rings is not determined by the ILR, but by
the centrifugal barrier that the gas driven inward by the bar torque
cannot overcome.

\begin{figure*}
\hspace{0.5cm}\includegraphics[angle=0, width=17cm]{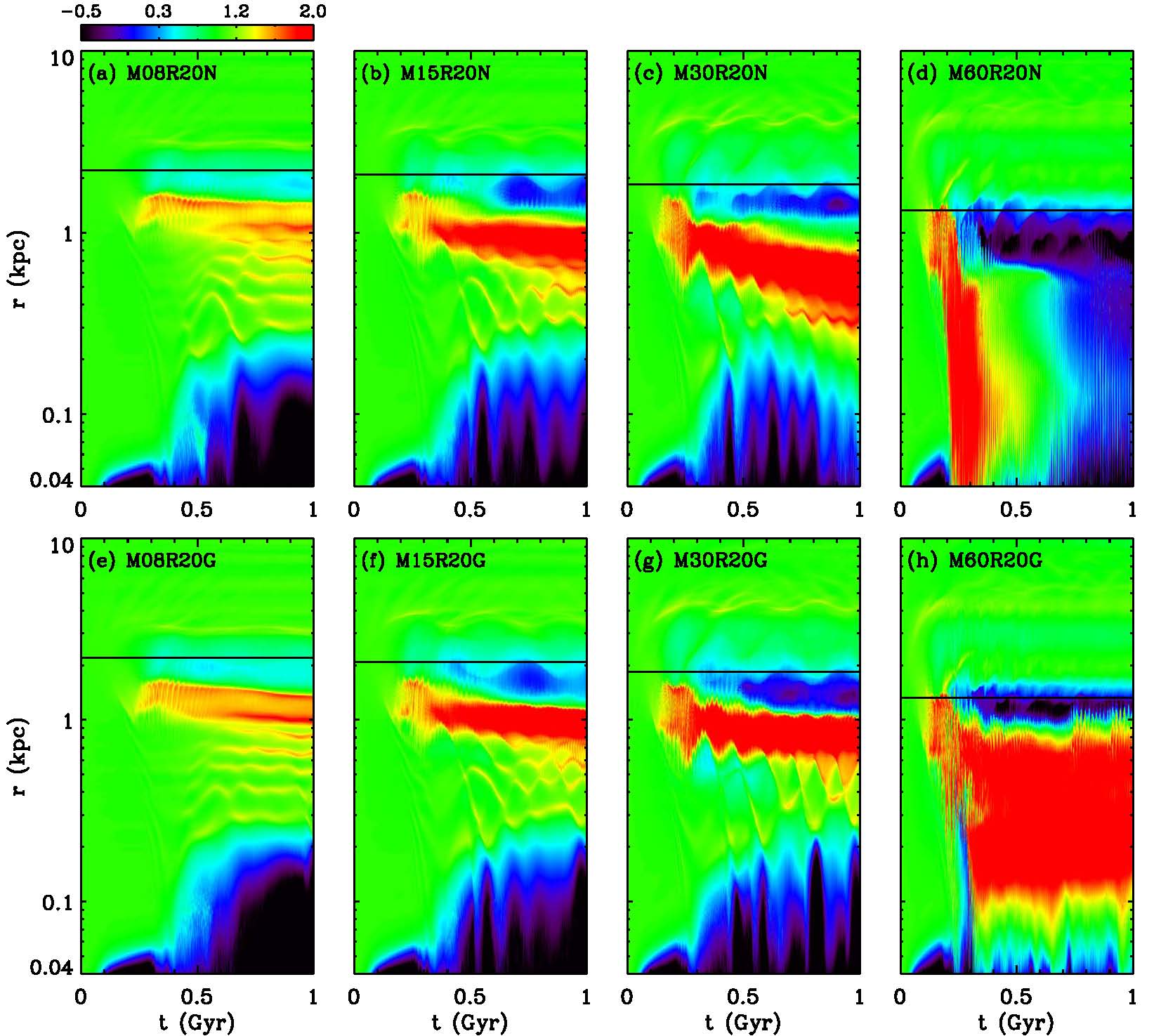}
\caption{Temporal changes of the azimuthally-averaged surface
density, $\langle\Sigma\rangle$, in logarithmic scale for all models
with $\R=2.0$. The upper row is for the non-self-gravitating models,
while the lower low is for the self-gravitating counterparts. The
horizontal line in each panel represents the location of the ILR.
Color bar labels
$\log(\langle\Sigma\rangle/\Sigma_0)$.\label{fig:ring_evol}}
\end{figure*}

\begin{figure}
\hspace{0.2cm}\includegraphics[angle=0, width=8.5cm]{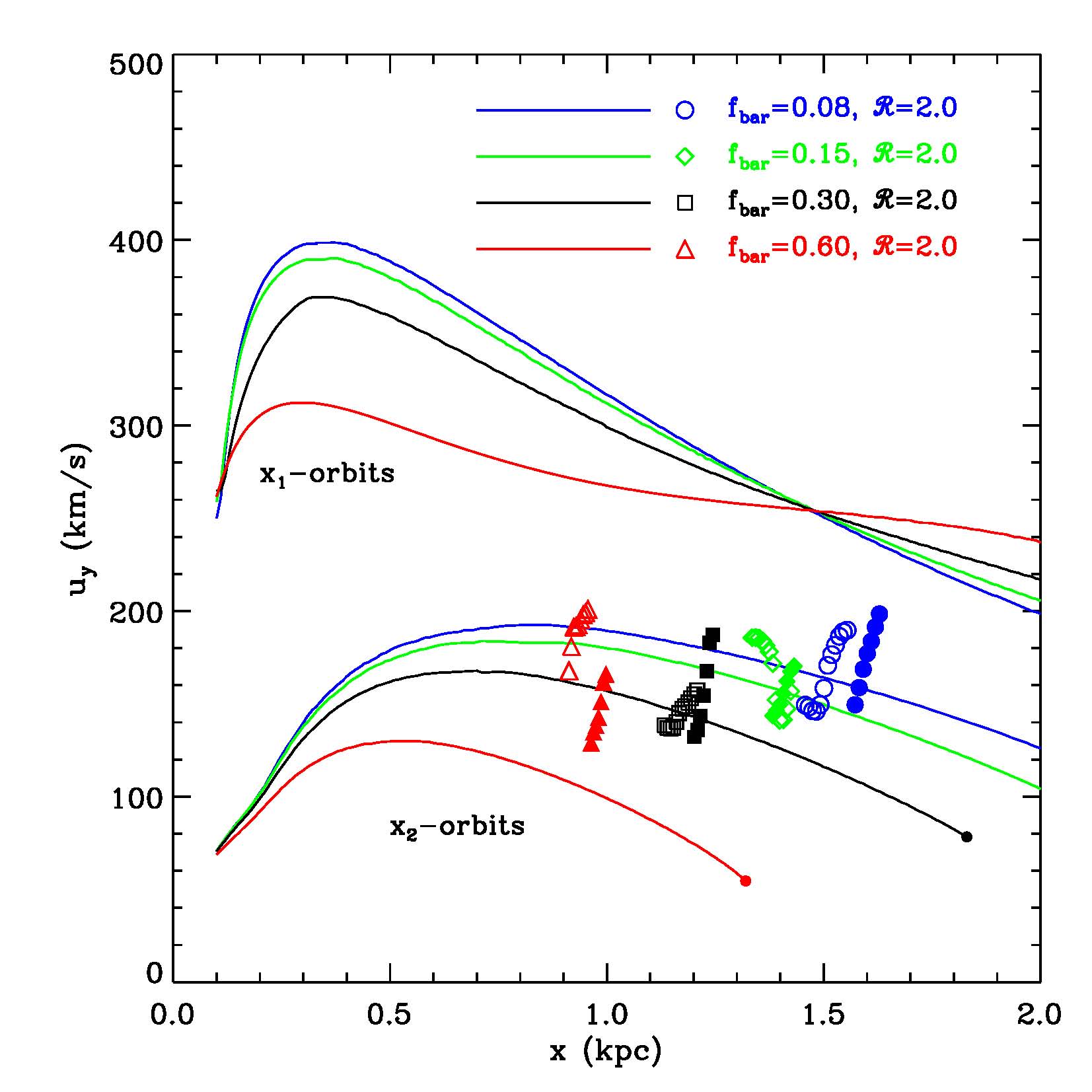}
\caption{Distributions of $y$-velocities in the rotating frame of
$\xone$-orbits (upper curves) and $\xtwo$-orbits (lower curves) at
$y=0$ as functions of $x$. Various symbols plot the numerical data
of the nuclear ring at $y=0$ when it begins to form at $t=0.25\Gyr$
for Model M08R20 (circles), $t=0.20\Gyr$ for Model M15R20
(triangles), $t=0.15\Gyr$ for Model M30R20, and $t=0.12\Gyr$ for
Model M60R20. Open and filled symbols give the results for the
non-self-gravitating and self-gravitating models, respectively. A
small dot at the tip of each of the lowest two curves denotes the
outermost $\xtwo$-orbit in that model. \label{fig:x12_vel}}
\end{figure}

The upper panels of Figure \ref{fig:ring_evol} show the temporal and
radial variation of the azimuthally-averaged surface density for all
non-self-gravitating models with $\R=2$, with the horizontal line
indicating the ILR in each model. Various open symbols in Figure
\ref{fig:x12_vel} plot the velocities $\uy$, seen in the rotating
frame with the bar, of the ring gas at $y=0$ when the ring is
beginning to form in the $\R=2$ models without self-gravity.  Also
plotted are the $y$-velocities of $\xone$-orbits ($\vyone$; upper
curves) and $\xtwo$-orbits ($\vytwo$; lower curves) when they cut
the $x$-axis. For fixed $\R$, a model with larger $\fbar$ has a
shallower external potential $\Ptot$ and thus a lower value of
$\vyone$ and $\vytwo$ at the same $x$. A small solid circle at the
tip of each of the lowest two curves marks the outermost
$\xtwo$-orbit beyond which no $\xtwo$-orbit exists in that model.

As mentioned before, the ring in Model M08R20N with $\fbar=0.08$
begins to form at $t\sim0.25\Gyr$. Already at this time, the ring
material has a velocity very similar to that of an $\xtwo$-orbit at
the ring position (open circles in Fig.\ \ref{fig:x12_vel}). Since
the bar torque is quite weak, the angular momentum of the gas that
is subsequently added to the ring is not much different from that of
the gas already in the ring, keeping $\Rring$ almost unchanged with
time (Fig.\ \ref{fig:ring_evol}a). As $\fbar$ (or $\Qb$) increases,
rings form earlier and locate closer to the center. For instance,
the ring in Model M30R20N with $\fbar=0.3$ begins to form at
$\Rring\sim 1.2\kpc$ when $t\sim0.15\Gyr$, with $u_y\simeq \vytwo
\ll \vyone$ (open squares in Fig.\ \ref{fig:x12_vel}).  With a
considerable loss of angular momentum at the dust-lane shocks, the
gas newly added to the ring has increasingly lower angular momentum,
causing it to shrink with time.   In addition, thermal pressure
perturbations make the orbits of ring material deviate from
$\xtwo$-orbits.  Some gases at the inner parts of the ring even take
on $\xone$-orbits.  This causes the ring not only to be distributed
more widely but also to shrink with time as it loses further angular
momentum due to supersonic collisions of the gases on $\xone$- and
$\xtwo$-orbits. Yet, the decreasing rate of the ring radius is still
quite small at $d\ln\Rring/dt\sim - 0.1\;\rm Gyr^{-1}$ in Model
M15R20N and $\sim - 0.4\;\rm Gyr^{-1}$ in Model M30R20N. In these
models, rings are still of the $\xtwo$ type (Fig.\
\ref{fig:inner}b).

In Model M60R20N with $\fbar=0.6$, on the other hand,  a large bar
torque takes the inflowing gas to $\Rring\sim0.9\kpc$ with velocity
$\uy\sim 190\kms$, which is just in between $\vyone= 272\kms$ and
$\vytwo=109\kms$ at that location (open triangles in Fig.\
\ref{fig:x12_vel}). Since $\uy \gg \vytwo$, the inflowing gas is
unable to settle on an $\xtwo$-orbit: the ring gas instead follows a
hybrid orbit that is inclined by about $\pring\sim17^\circ$ with
respect to the $x$-axis at $t\sim0.12\Gyr$. With a subsequent
addition of low angular-momentum gas from outside, the ring shrinks
and becomes more eccentric and inclined with time. The ring position
angle is increased to $\pring\sim40^\circ$ at $t\sim 0.2\Gyr$ when
the ring becomes very eccentric to touch the inner boundary and
starts to lose a significant amount of its mass through the inner
boundary (Fig.\ \ref{fig:ring_evol}d). At the same time, the
inclined ring precesses slowly in the clockwise direction due to the
bar torque that tends to align the ring parallel to its major axis
(see \citealt{ks12} for the precession of a ring in magnetized
models). Figure \ref{fig:inner}c shows that the position angle of
the ring in Model M60R20N is $\pring\sim65^\circ$ at $t=0.3\Gyr$,
which eventually becomes $90^\circ$ at $t\sim0.38\Gyr$ (i.e.,
$\xone$-type ring).

The dominance (or absence) of the $\xone$ family of the gaseous
orbits in our non-self-gravitating simulations appears to be
determined not only by the amount of angular momentum loss at
dust-lane shocks but also by the kinetic energies of $\xone$- and
$\xtwo$-orbits allowed under a given external potential. The former
is measured by $\Qb$, while the latter is affected by $\R$. For
example, Figure \ref{fig:all_no}s shows that Model M30R35N with
$\Qb=0.39$ still possesses a well-defined $\xtwo$-type ring.  In
this model, $\Qb$ is large enough to decrease the velocity of the
ring material, while $\R$ (together with $\fbar$) sets $\xone$- and
$\xtwo$-orbits such that the rotational velocity of the ring
material is larger than $\vytwo$ only slightly, but much less than
$\vyone$, leading to an $\xtwo$-type ring. With an even larger value
of $\Qb=0.67$, the ring in Model M60R35N has $\uy\sim120\kms$ at
$r\sim0.8\kpc$ when it begins to shape at $t=0.08\Gyr$. However,
this model has the outermost $\xtwo$-orbit at $x_c=0.43\kpc$, so
that the ring material has no choice but to follow $\xone$-orbits.
The ring subsequently becomes smaller and more eccentric as low
angular-momentum gas is continuously added. As the ring material
directly plunges through the inner boundary, it rapidly decays at a
rate $d\ln\Rring/dt\sim - 15 \;\rm Gyr^{-1}$. In this model,  most
of the ring gas is accreted to the center by $t\sim0.25\Gyr$, with a
small quantity of gas in filamentary shapes moving about the center
along $\xone$- rather than $\xtwo$-orbits.

\subsubsection{Effects of Self-gravity}

Nuclear rings form typically at $r\sim0.5-1.5\kpc$. Equation
(\ref{eq:QT}) suggests that a ring could possibly be gravitationally
unstable if its peak density exceeds $\sim500\Surf$.\footnote{If the
density enhancement occurs in an angular-momentum-conserving
fashion, $\kappa\propto\Sigma^{1/2}$ and $Q_T\propto\Sigma^{-1/2}$
as in the case of spiral shocks (e.g., \citealt{bal85,kim02}). Rings
in barred galaxies form by gathering material that \emph{lost}
angular momentum by a non-axisymmetirc bar torque as well as shocks.
In our models, we found $\kappa$ calculated from the
azimuthally-averaged rotational velocity is similar to its initial
profile, suggesting that density compression in the ring does not
alter $\kappa$ much.} This happens in most of our self-gravitating
models except for Models M08R15G and M15R15G where the bar torque
with $\Qb\simlt0.04$ is too weak to cause significant gas inflows to
the central regions.  If a ring were uniform in models with
$\Qb\simgt0.05$, it would have been prone to gravitational
fragmentation.  However, rings are already clumpy even in
non-self-gravitating models as a result of the wiggle instability of
dust-lane shocks.   In models with intermediate bar torque (i.e,
$\Qb\sim0.1-0.2$), therefore, clumps in nuclear rings simply acquire
more mass and become denser due to self-gravity.  In models with
$\Qb\simgt0.2$, self-gravity produces additional fragments that make
rings more clumpy in comparisons with non-self-gravitating
counterparts (see, e.g., Figs.\ \ref{fig:all_no}k and
\ref{fig:all_sg}k). The typical mass of clumps produced is
$\sim10^6-10^7\Msun$. We note that the maximum density such clumps
can attain is limited by numerical resolution. Self-gravitating
clouds are resolved only when $\Delta r < \lambda_J/4$
\citep{tru97,tru98}, where $\Delta r$ is the grid spacing and
$\lambda_J = c_s^2/(G\Sigma)$ is the local Jeans length. Therefore,
clumps with $\Sigma\simgt 10^3\Surf$ at $r\sim1\kpc$ are
gravitationally unresolved in our simulations.

Another important effect of self-gravity is that it makes nuclear
rings larger compared to those in non-self-gravitating models.
Figures \ref{fig:inner} and \ref{fig:ring_evol} show that compared
to in Model M30R20N, the ring in Model M30R20G is larger in size at
$t=0.3\Gyr$ and decays less afterward (see also Fig.\
\ref{fig:t800}). This is because self-gravity deepens the total
gravitational potential at the location of a ring, making the orbits
of the ring material relatively intact and resistant to external
perturbations that tend to change the orbits. Therefore, the radial
decay of a ring due to the addition of low angular-momentum gas from
outside is slower in self-gravitating models.  The effect of thermal
pressure in dispersing the ring material spatially is also smaller,
resulting in a narrower and larger ring than in non-self-gravitating
models.

Finally, we discuss rings in self-gravitating models with a massive
bar ($\fbar=0.6$). Figure \ref{fig:inner} shows that Model M60R20G
possesses $\xtwo$-type, double rings at $t=0.3\Gyr$, which is unlike
Model M60R20N that has an inclined, single ring. As explained above,
a large bar torque in Model M60R20N produces an inclined ring at
$t\sim0.12\Gyr$ that processes with time to become of the
$\xone$-type.  Strong self-gravity in Model M60R20G prevents the
precession of the ring by providing additional non-axisymmetric
torque, gradually aligning its long axis parallel to the bar minor
axis. At $t=0.22\Gyr$, the ring is still quite eccentric and forms
many high-density clumps via gravitational instability. When dense
(unresolved) clumps move close to the galaxy center on their orbits
($t\sim0.26\Gyr$), they gravitationally interact with other clumps
at the opposite side. These interacting clumps lower their orbits to
smaller $r$, gather gas from the ring outside, and form a inner ring
at $t=0.29\Gyr$.  The inner ring slowly dissipates as it loses mass
through the inner boundary. In Model M60R25G, self-gravity results
in an $\xtwo$-type ring. But, the ring in this model is already so
small that the gravitational interaction of dense clumps do not
produce an obvious inner ring. Small, $\xtwo$-type rings in Models
M60R30G and M60R35G (Fig.\ \ref{fig:all_sg}p,t) become weaker with
time and vanish at $t\sim0.75\Gyr$, so that they are not considered
as being permanent (Fig.\ \ref{fig:t800}p).

\begin{figure*}
\hspace{0.5cm}\includegraphics[angle=0, width=17cm]{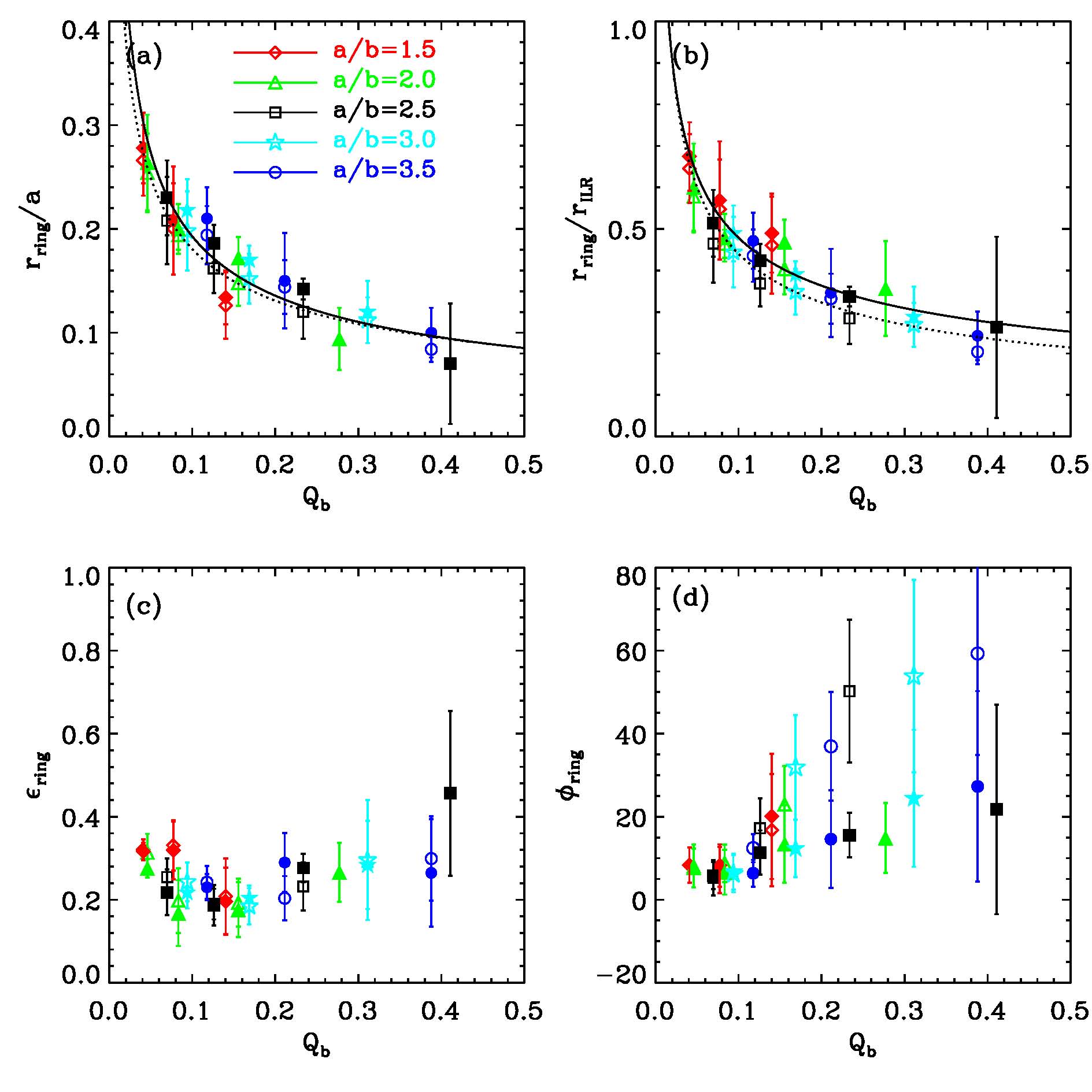}
\caption{Dependence on the bar strength $\Qb$ of the ring size
$\Rring$ relative (a) to the bar semimajor axis $a$ and (b) to the
ILR radius $\ILR$, (c) the ring ellipticity $\ering$, and (d) the
ring position angle $\pring$ for models with $\xtwo$-type rings. In
each panel, symbols give the mean values averaged over
$t=0.3$--$1.0\Gyr$, while errorbars represent the standard
deviations.  Open and filled symbols are for non-self-gravitating
and self-gravitating models, respectively. Dotted and solid lines in
(a) and (b) draw equations (\ref{eq:Rring_a}) and
(\ref{eq:Rring_ILR}) for non-self-gravitating and self-gravitating
models, respectively. \label{fig:ring}}
\end{figure*}

\subsubsection{Ring Properties}

To quantify the ring properties, we at each time calculate the mean
density $\Sring$, mean radius $\Rring$, ellipticity
$\ering=1-b_r/a_r$ with $a_r$ and $b_r$ denoting the semi-major and
minor axes, respectively, and the position angle $\pring$ of the
long axis of the ring, and take their temporal averages over
$t=0.3$--$1.0\Gyr$. The resulting mean values and standard
deviations are given in Tables \ref{tbl:ring_no} and
\ref{tbl:ring_sg} for non-self-gravitating and self-gravitating
models, respectively. Figure \ref{fig:ring} plots these as various
symbols and errorbars as functions of $\Qb$ for models with
$\xtwo$-type rings.  Open symbols are for the non-self-gravitating
models, while the self-gravitating results are plotted as filled
symbols. It turns out that $\Rring$ is best fitted solely by $\Qb$
not in combination with $\R$.  Our best fits of $\Rring$ against
$\Qb$ are
\begin{equation}\label{eq:Rring_a}
\frac{\Rring}{a} = \left\{\begin{array}{ll}
 0.062\Qb^{-0.46}, & \textrm{without self-gravity}, \\
 0.059\Qb^{-0.51}, & \textrm{with self-gravity},
\end{array}\right.
\end{equation}
in terms of the bar major axis, and
\begin{equation}\label{eq:Rring_ILR}
\frac{\Rring}{\ILR} = \left\{\begin{array}{ll}
 0.16\Qb^{-0.45}, & \textrm{without self-gravity}, \\
 0.19\Qb^{-0.40}, & \textrm{with self-gravity},
\end{array}\right.
\end{equation}
in terms of the ILR radius.

\begin{deluxetable}{lcccc}
\tabletypesize{\footnotesize} \tablewidth{0pt}
\tablecaption{Properties of Nuclear Rings in Non-self-gravitating
Models\label{tbl:ring_no}} \tablehead{ \colhead{Model}            &
\colhead{$\Sring/\Sigma_0$}& \colhead{$\Rring$ (kpc)}   &
\colhead{$\ering$}         & \colhead{$\pring$ (deg)} } \startdata
M08R15N &     $\cdots$     &      $\cdots$    &      $\cdots$    &   $\cdots$       \\
M08R20N &  1.5  $\pm$  0.2 & 1.27  $\pm$ 0.19 & 0.31  $\pm$ 0.05 &  7.6  $\pm$  4.7 \\
M08R25N &  4.1  $\pm$  1.5 & 1.04  $\pm$ 0.21 & 0.25  $\pm$ 0.05 &  5.3  $\pm$  4.3 \\
M08R30N &  5.1  $\pm$  1.8 & 0.99  $\pm$ 0.19 & 0.24  $\pm$ 0.05 &  6.2  $\pm$  4.4 \\
M08R35N &  4.6  $\pm$  2.0 & 0.97  $\pm$ 0.14 & 0.24  $\pm$ 0.04 & 12.4  $\pm$  3.4 \\
\hline
M15R15N &  1.5  $\pm$  0.3 & 1.33  $\pm$ 0.17 & 0.32  $\pm$ 0.02 &  8.3  $\pm$  4.3 \\
M15R20N &  3.6  $\pm$  1.4 & 0.97  $\pm$ 0.07 & 0.20  $\pm$ 0.08 &  8.8  $\pm$  4.6 \\
M15R25N &  9.3  $\pm$  4.4 & 0.81  $\pm$ 0.12 & 0.19  $\pm$ 0.05 & 17.2  $\pm$  7.2 \\
M15R30N &  9.8  $\pm$  3.9 & 0.76  $\pm$ 0.12 & 0.18  $\pm$ 0.04 & 31.9  $\pm$ 12.6 \\
M15R35N & 14.1  $\pm$  9.8 & 0.72  $\pm$ 0.13 & 0.20  $\pm$ 0.05 & 36.9  $\pm$ 13.1 \\
\hline
M30R15N &  2.5  $\pm$  0.8 & 1.00  $\pm$ 0.22 & 0.33  $\pm$ 0.06 &  7.1  $\pm$  5.6 \\
M30R20N &  8.2  $\pm$  4.4 & 0.74  $\pm$ 0.11 & 0.19  $\pm$ 0.06 & 22.9  $\pm$  9.3 \\
M30R25N & 22.9  $\pm$ 13.0 & 0.60  $\pm$ 0.13 & 0.23  $\pm$ 0.06 & 50.2  $\pm$ 17.2 \\
M30R30N & 28.9  $\pm$ 14.0 & 0.56  $\pm$ 0.11 & 0.30  $\pm$ 0.14 & 53.8  $\pm$ 23.2 \\
M30R35N & 38.0  $\pm$ 12.4 & 0.42  $\pm$ 0.06 & 0.30  $\pm$ 0.10 & 59.3  $\pm$ 24.5 \\
\hline
M60R15N & 4.6  $\pm$  2.5 & 0.63  $\pm$ 0.16 & 0.21  $\pm$ 0.09 & 16.8  $\pm$ 13.5 \\
M60R20N & 1.4  $\pm$  1.2 & 0.15  $\pm$ 0.10 & 0.63  $\pm$ 0.22 & 36.0  $\pm$ 36.7 \\
M60R25N & 0.8  $\pm$  2.0 & 0.21  $\pm$ 0.14 & 0.47  $\pm$ 0.17 & 87.8  $\pm$ 17.3 \\
M60R30N &  $\cdots$  &      $\cdots$   &      $\cdots$   &   $\cdots$     \\
M60R35N &  $\cdots$  &      $\cdots$   &      $\cdots$   &
$\cdots$
\enddata
\tablecomments{$\Sring$, $\Rring$, $\ering$ and $\pring$ denote the
mean density, mean radius, ellipticity, and position angle of the
long axis of a ring, respectively.}
\end{deluxetable}

Equations (\ref{eq:Rring_a}) and (\ref{eq:Rring_ILR}) are plotted in
Figure \ref{fig:ring}a,b as dotted and solid lines for
non-self-gravitating and self-gravitating models, respectively.
Overall, rings in self-gravitating models are about $\sim5$--20\%
larger than in non-self-gravitating models. The decrease of $\Rring$
with $\Qb$ is of course due to the fact that a stronger bar torque
drives the gas closer to the galaxy center by removing larger
angular momentum from it.  Note that $\Rring/\ILR$ is less than
unity and can be as small as $\sim0.2$, demonstrating again that the
ring position is not determined by the ILR but by $\Qb$.  We will
show in Section \ref{sec:sum} that the decreasing tendency of
$\Rring$ with $\Qb$ is qualitatively consistent with the
observational results of \citet{com09} who found that nuclear rings
are smaller in more strongly barred galaxies. In our models, nuclear
rings are in general elliptical with $\ering\sim0.2$--$0.3$,
insensitive to $\Qb$ and self-gravity. They are well aligned with
the bar minor axis when $\Qb <0.1$, while becoming more inclined as
$\Qb$ increases.

\subsection{Nuclear Spirals}\label{sec:nsp}

There are two types of perturbations that can excite nuclear spirals
in the central regions: the non-axisymmetric bar potential and sonic
perturbations from a nuclear ring.  At early time before a ring
forms, the bar potential is a lone perturbing agent of $m=2$ spiral
waves. As explained in Paper I, our galaxy models with a central BH
have $d(\Omega-\kappa/2)/dr>0$ at $\Rmin \leq r\leq \Rmax$ and
$d(\Omega-\kappa/2)/dr<0$ otherwise, with the local minimum and
maximum of the $\Omega-\kappa/2$ curve occurring at
$\Rmin\sim0.2\kpc$ and $\Rmax\sim0.4$--$0.6\kpc$ depending on
$\fbar$ and $\R$. Thus, the spiral waves at $r<\Rmin$ are trailing,
while they are leading at $\Rmin \leq r\leq \Rmax$ early time. Since
the bar potential is nearly axisymmetric near the center, these
waves would remain weak unless additional perturbations are
supplied. In models with an $\xtwo$-type ring, a ring that forms
outside the spirals provides sonic perturbations that propagate
inward in the form of trailing waves. These waves interact
constructively (destructively) with the trailing (leading) part of
the spirals.  Consequently, the outer leading parts are destroyed,
while the inner trailing spirals grow stronger and extend outward to
make contact with the ring (Fig.\ \ref{fig:inner}a,b). On the other
hand, models with an $\xone$-type ring do not retain nuclear spirals
since the ring gas on eccentric orbits wipes out inner coherent
spiral structures (Fig.\ \ref{fig:inner}c). Even though Model
M30R35N has an $\xtwo$-type ring, it is so small that strong
perturbations from it inhibit the growth of spirals inside.
Therefore, the presence of nuclear spirals requires two conditions:
they should be of the $\xtwo$ type and sufficiently large (perhaps
larger than $\Rmax$).

\begin{deluxetable}{lcccc}
\tabletypesize{\footnotesize} \tablewidth{0pt}
\tablecaption{Properties of Nuclear Rings in Self-gravitating
Models\label{tbl:ring_sg}} \tablehead{ \colhead{Model}            &
\colhead{$\Sring/\Sigma_0$}& \colhead{$\Rring$ (kpc)}   &
\colhead{$\ering$}         & \colhead{$\pring$ (deg)} } \startdata
M08R15G &     $\cdots$  &      $\cdots$   &      $\cdots$   &   $\cdots$     \\
M08R20G &   2.4  $\pm$   1.0 & 1.32  $\pm$ 0.23 & 0.28  $\pm$ 0.02 &  8.2  $\pm$  5.1 \\
M08R25G &   4.2  $\pm$   1.5 & 1.15  $\pm$ 0.18 & 0.22  $\pm$ 0.05 &  5.8  $\pm$  3.3 \\
M08R30G &   5.6  $\pm$   1.9 & 1.09  $\pm$ 0.15 & 0.22  $\pm$ 0.04 &  6.7  $\pm$  4.3 \\
M08R35G &   6.1  $\pm$   2.2 & 1.05  $\pm$ 0.15 & 0.25  $\pm$ 0.03 &  6.4  $\pm$  3.2 \\
\hline
M15R15G &   1.6  $\pm$   0.4 & 1.39  $\pm$ 0.17 & 0.32  $\pm$ 0.02 &  8.3  $\pm$  4.2 \\
M15R20G &   5.6  $\pm$   2.3 & 1.00  $\pm$ 0.12 & 0.17  $\pm$ 0.08 &  6.6  $\pm$  5.4 \\
M15R25G &   9.2  $\pm$   4.6 & 0.93  $\pm$ 0.09 & 0.19  $\pm$ 0.04 & 11.3  $\pm$  5.3 \\
M15R30G &  12.6  $\pm$   6.0 & 0.85  $\pm$ 0.07 & 0.20  $\pm$ 0.03 & 12.3  $\pm$  6.8 \\
M15R35G & 111.5  $\pm$  56.7 & 0.75  $\pm$ 0.23 & 0.29  $\pm$ 0.07 & 14.6  $\pm$ 11.8 \\
\hline
M30R15G &   2.8  $\pm$   1.1 & 1.04  $\pm$ 0.26 & 0.32  $\pm$ 0.07 &  8.2  $\pm$  5.1 \\
M30R20G &  11.2  $\pm$   5.0 & 0.86  $\pm$ 0.10 & 0.18  $\pm$ 0.07 & 13.4  $\pm$  9.3 \\
M30R25G &  18.4  $\pm$   7.5 & 0.71  $\pm$ 0.05 & 0.28  $\pm$ 0.03 & 15.6  $\pm$  5.4 \\
M30R30G &  81.2  $\pm$  23.4 & 0.60  $\pm$ 0.15 & 0.28  $\pm$ 0.11 & 24.4  $\pm$ 16.5 \\
M30R35G &  78.7  $\pm$  17.1 & 0.50  $\pm$ 0.12 & 0.26  $\pm$ 0.13 & 27.3  $\pm$ 23.0 \\
\hline
M60R15G &   5.2  $\pm$   2.5 & 0.67  $\pm$ 0.13 & 0.20  $\pm$ 0.08 & 20.1  $\pm$ 15.1 \\
M60R20G &  32.8  $\pm$  10.0 & 0.47  $\pm$ 0.15 & 0.27  $\pm$ 0.07 & 14.8  $\pm$  8.4 \\
M60R25G &  69.5  $\pm$  10.5 & 0.35  $\pm$ 0.29 & 0.46  $\pm$ 0.20 & 21.8  $\pm$ 25.3 \\
M60R30G &      $\cdots$  &      $\cdots$   &      $\cdots$   &   $\cdots$     \\
M60R35G &      $\cdots$  &      $\cdots$   &      $\cdots$   &
$\cdots$
\enddata
\tablecomments{$\Sring$, $\Rring$, $\ering$ and $\pring$ denote the
mean density, mean radius, ellipticity, and position angle of the
long axis of a ring, respectively.}
\end{deluxetable}

\begin{deluxetable*}{lll|lll}
\tabletypesize{\footnotesize} \tablewidth{0pt}
\tablecaption{Time-averaged Mass Inflow Rate and Its
Dispersion\label{tbl:mdot}} \tablehead{ \colhead{Model} &
\colhead{$\Mdot$ } & \colhead{$\Delta \Mdot$}  & \colhead{Model} &
\colhead{$\Mdot$} & \colhead{$\Delta \Mdot$}
\\
\colhead{ } & \colhead{$(\Msun\;\yr^{-1})$ } &
\colhead{$(\Msun\;\yr^{-1})$ } & \colhead{ } &
\colhead{$(\Msun\;\yr^{-1})$ } & \colhead{$(\Msun\;\yr^{-1})$ } }
\startdata M08R15N & $2.4\times 10^{-3}$ & $1.7\times 10^{-3}$ &
M08R15G & $2.2\times 10^{-3}$ & $1.6\times 10^{-3}$ \\
M08R20N & $2.3\times 10^{-3}$ & $2.1\times 10^{-3}$ &
M08R20G & $2.1\times 10^{-3}$ & $1.9\times 10^{-3}$ \\
M08R25N & $2.9\times 10^{-3}$ & $1.9\times 10^{-3}$ &
M08R25G & $2.5\times 10^{-3}$ & $1.7\times 10^{-3}$ \\
M08R30N & $3.1\times 10^{-3}$ & $2.1\times 10^{-3}$ &
M08R30G & $2.4\times 10^{-3}$ & $1.5\times 10^{-3}$ \\
M08R35N & $3.2\times 10^{-3}$ & $2.0\times 10^{-3}$ &
M08R35G & $2.4\times 10^{-3}$ & $1.3\times 10^{-3}$ \\
\hline M15R15N & $2.3\times 10^{-3}$ & $2.1\times 10^{-3}$ &
M15R15G & $2.2\times 10^{-3}$ & $2.0\times 10^{-3}$ \\
M15R20N & $3.2\times 10^{-3}$ & $2.2\times 10^{-3}$ &
M15R20G & $2.4\times 10^{-3}$ & $1.8\times 10^{-3}$ \\
M15R25N & $3.3\times 10^{-3}$ & $2.3\times 10^{-3}$ &
M15R25G & $3.0\times 10^{-3}$ & $1.9\times 10^{-3}$ \\
M15R30N & $2.6\times 10^{-3}$ & $1.7\times 10^{-3}$ &
M15R30G & $2.6\times 10^{-3}$ & $2.3\times 10^{-3}$ \\
M15R35N & $2.5\times 10^{-3}$ & $2.3\times 10^{-3}$ &
M15R35G & $1.2\times 10^{-1}$ & $7.7\times 10^{ 0}$ \\
\hline M30R15N & $3.1\times 10^{-3}$ & $2.7\times 10^{-3}$ &
M30R15G & $2.7\times 10^{-3}$ & $2.2\times 10^{-3}$ \\
M30R20N & $5.2\times 10^{-3}$ & $3.9\times 10^{-3}$ &
M30R20G & $2.9\times 10^{-3}$ & $2.3\times 10^{-3}$ \\
M30R25N & $2.0\times 10^{-3}$ & $3.2\times 10^{-3}$ &
M30R25G & $4.2\times 10^{-3}$ & $2.9\times 10^{-3}$ \\
M30R30N & $2.5\times 10^{-2}$ & $5.4\times 10^{-2}$ &
M30R30G & $3.6\times 10^{-1}$ & $2.0\times 10^{ 1}$ \\
M30R35N & $1.7\times 10^{-1}$ & $2.3\times 10^{-1}$ &
M30R35G & $3.8\times 10^{-1}$ & $1.9\times 10^{ 1}$ \\
\hline M60R15N & $7.5\times 10^{-3}$ & $8.6\times 10^{-3}$ &
M60R15G & $3.9\times 10^{-3}$ & $4.6\times 10^{-3}$ \\
M60R20N & $3.2\times 10^{-1}$ & $8.1\times 10^{-1}$ &
M60R20G & $1.2\times 10^{-2}$ & $2.6\times 10^{-1}$ \\
M60R25N & $5.4\times 10^{-1}$ & $2.3\times 10^{ 0}$ &
M60R25G & $6.3\times 10^{-1}$ & $4.3\times 10^{ 1}$ \\
M60R30N & $8.8\times 10^{-1}$ & $2.2\times 10^{ 0}$ &
M60R30G & $7.0\times 10^{-1}$ & $3.2\times 10^{ 1}$ \\
M60R35N & $9.2\times 10^{-1}$ & $2.0\times 10^{ 0}$ & M60R35G &
$1.2\times 10^{ 0}$ & $5.5\times 10^{ 1}$
\enddata
\tablecomments{ Time average of $\Mdot$ is taken over
$t=0.1-1.0\Gyr$. }
\end{deluxetable*}

\begin{figure}
\hspace{0.5cm}\includegraphics[angle=0, width=8cm]{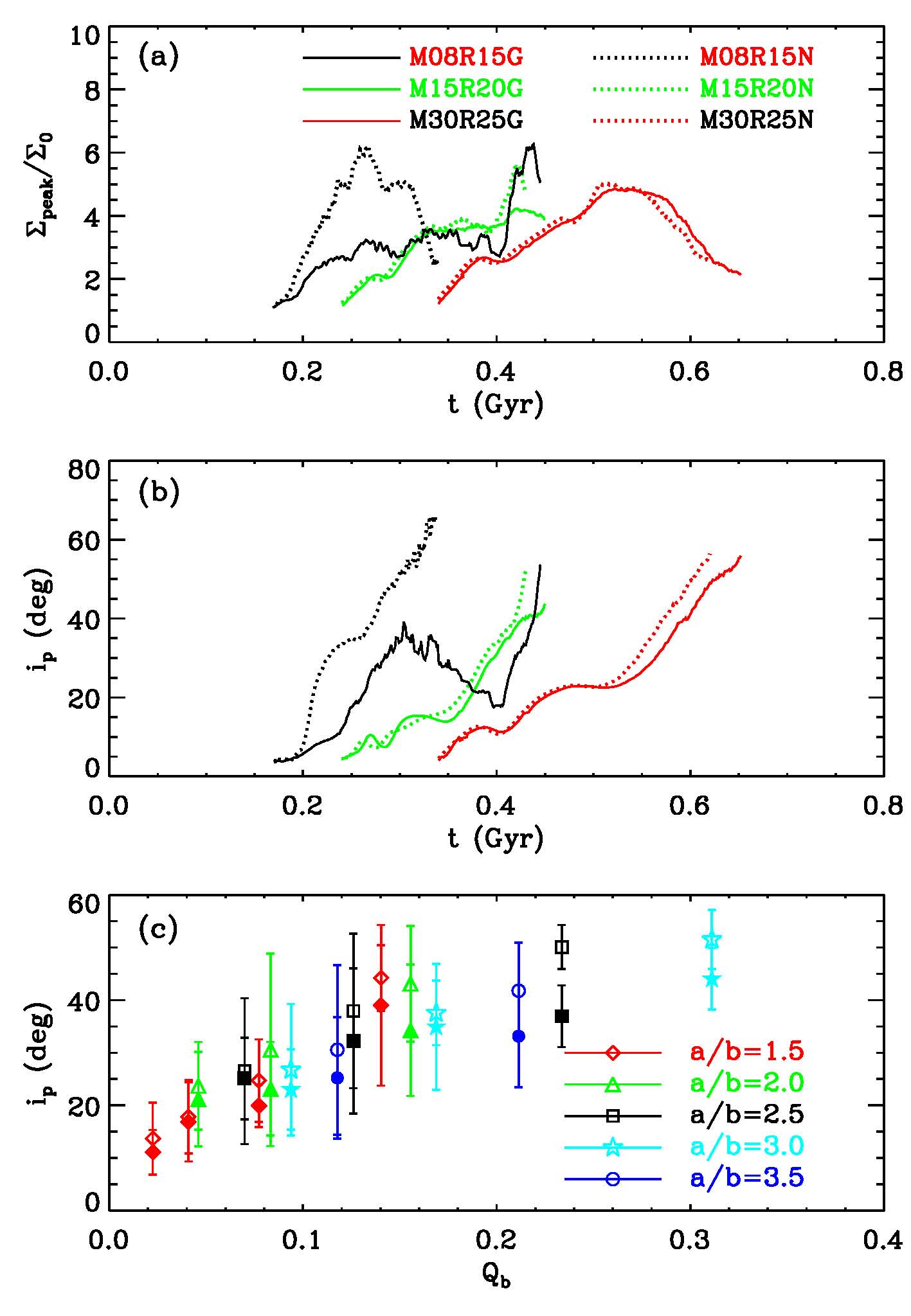}
\caption{Temporal evolution of (a) the peak surface density $\pSig$
and (b) the pitch angle $\ip$ of nuclear spirals at $r=0.2\kpc$ in
Models M08R15, M15R20, and M30R25. Dotted and solid lines are for
non-self-gravitating and self-gravitating models, respectively. (c)
Mean values (symbols) of $\ip$ averaged over $t=0.3$--$0.5\Gyr$ with
the standard deviations (errorbars) as a function of $\Qb$ for all
models with nuclear spirals. Open and filled symbols are for
non-self-gravitating and self-gravitating models, respectively.
\label{fig:nsp}}
\end{figure}

Unlike dust lanes and nuclear rings, both of which remain relatively
stationary after the bar potential is fully turned on, we find that
nuclear spirals do not achieve a quasi-steady state in that their
peak density $\pSig$ and pitch angle $\ip$ vary substantially with
time. Figure \ref{fig:nsp}a,b plots the temporal changes of $\pSig$
and $\ip$ measured at $r=0.2\kpc$ for a few selected models. Dotted
and solid lines correspond to non-self-gravitating and
self-gravitating models, respectively. Compared to the other
non-self-gravitating models, Model M30R25N forms nuclear spirals
earlier due to a larger bar torque. In general, nuclear spirals tend
to unwind (i.e., $\ip$ increases) as they become stronger. This
appears to be a generic consequence of the nonlinear effect.
\citet{lee99} showed analytically that dispersion relations of
nonlinear waves involve the wavenumber, frequency, and wave
amplitude all together, which is unlike in the linear-wave case
where the wavenumber and frequency are independent of the amplitude.
They further showed that the wavenumber of nonlinear spiral waves is
a decreasing function of the amplitude and that nonlinear trailing
waves unwind and become more nonlinear as they propagate inward due
to the increase in the angular momentum flux carried by the waves.
This nonlinear growth and unwinding of traveling waves is consistent
with the behavior of nuclear spirals formed in our models.

Figure \ref{fig:nsp} also shows that nuclear spirals in
self-gravitating models unwind more slowly than the
non-self-gravitating counterparts. This is presumably because
non-axisymmetric waves corotating with the bar in self-gravitating
models should have a larger radial wavenumber $k_r$ than those in
non-self-gravitating models in order to have the same frequency.
With larger $k_r$, the angle between the gas streamlines and nuclear
spirals are smaller in self-gravitating models. This leads to a
smaller departure of gas trajectories from the circular motions,
resulting in lower $\pSig$.

All nuclear spirals in our models are logarithmic in shape.  They
grow with time and eventually develop into shocks, with the shock
formation epoch delayed progressively with decreasing $\Qb$. For
instance, the spirals in Models M30R25N, M15R20N, and M08R15N become
shocks at $t\sim0.26$, $0.42$, and $0.50\Gyr$, respectively, after
which $\pSig$ decreases as most of the central gas inside the ring
is accreted through the inner boundary. The shock formation time of
nuclear spirals are delayed to $t\sim0.43$, $0.45$, and $0.55\Gyr$
for self-gravitating Models M30R25G, M15R20G, and M08R15G,
respectively. At a given time, therefore, the nuclear spirals in
smaller-$\Qb$ models tend to be more tightly wound, while those in
larger-$\Qb$ models tend to be more open and, sometimes, even
shocked (Fig.\ \ref{fig:inner}). This is illustrated quantitatively
in Figure \ref{fig:nsp}c where the temporal averages of $\ip$ and
the standard deviations over $t=0.3$--$0.5\Gyr$ are plotted for
models with appreciable spirals during this time interval. The
increasing behavior of $\ip$ with $\Qb$ is consistent with the
observational result of \citet{pee06} that nuclear spirals tend to
be tightly wound in weakly barred galaxies.

\subsection{Mass Inflow Rate}\label{sec:mdot}

\begin{figure*}
\hspace{0.5cm}\includegraphics[angle=0, width=17cm]{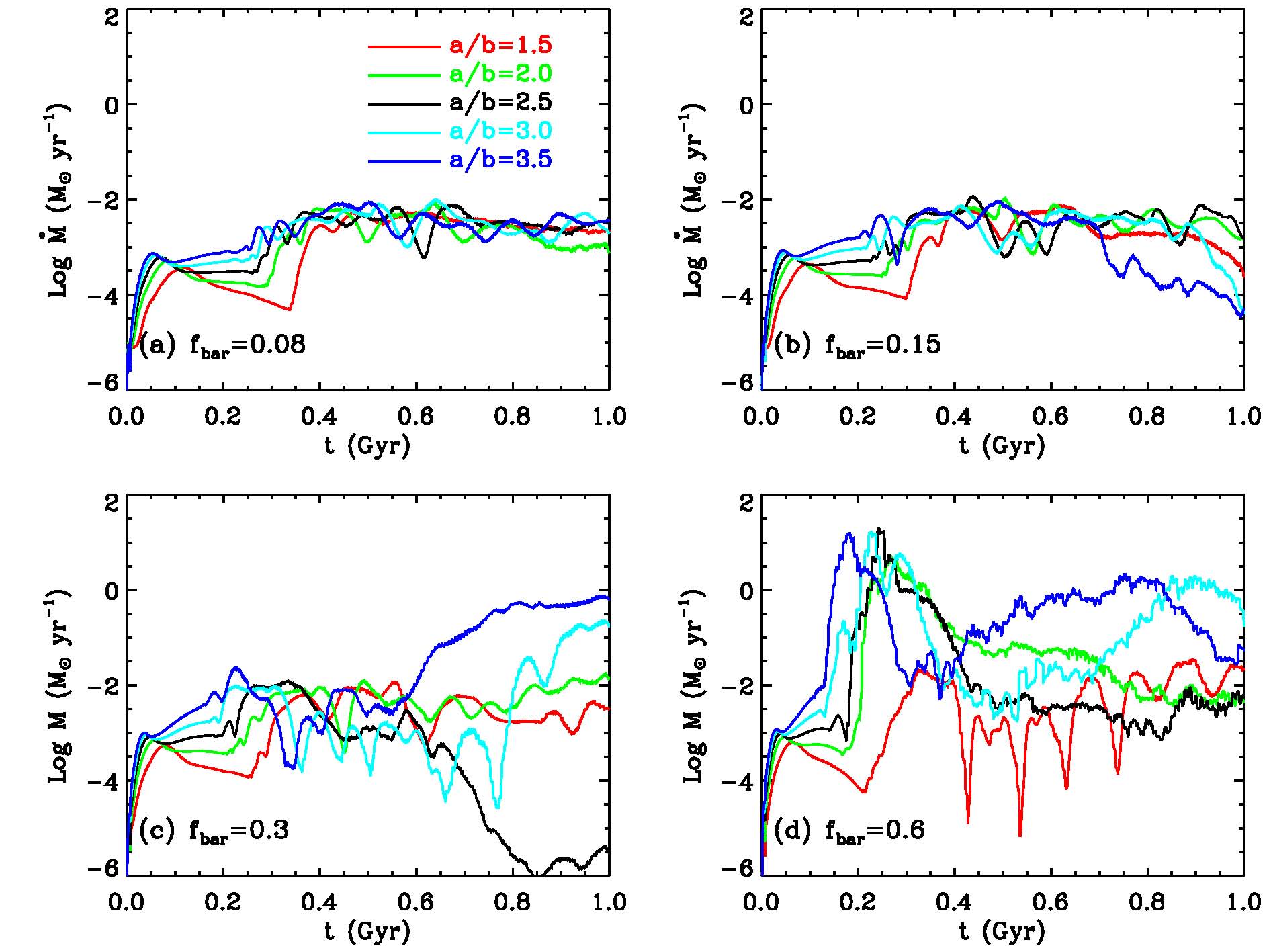}
\caption{ Temporal variations of the mass inflow rates $\Mdot$ for
non-self-gravitating models with (a) $\fbar=0.08$, (b) $\fbar=0.15$,
(c) $\fbar=0.3$, and (d) $\fbar=0.6$. \label{fig:mdot_no}}
\end{figure*}

\begin{figure*}
\hspace{0.5cm}\includegraphics[angle=0, width=17cm]{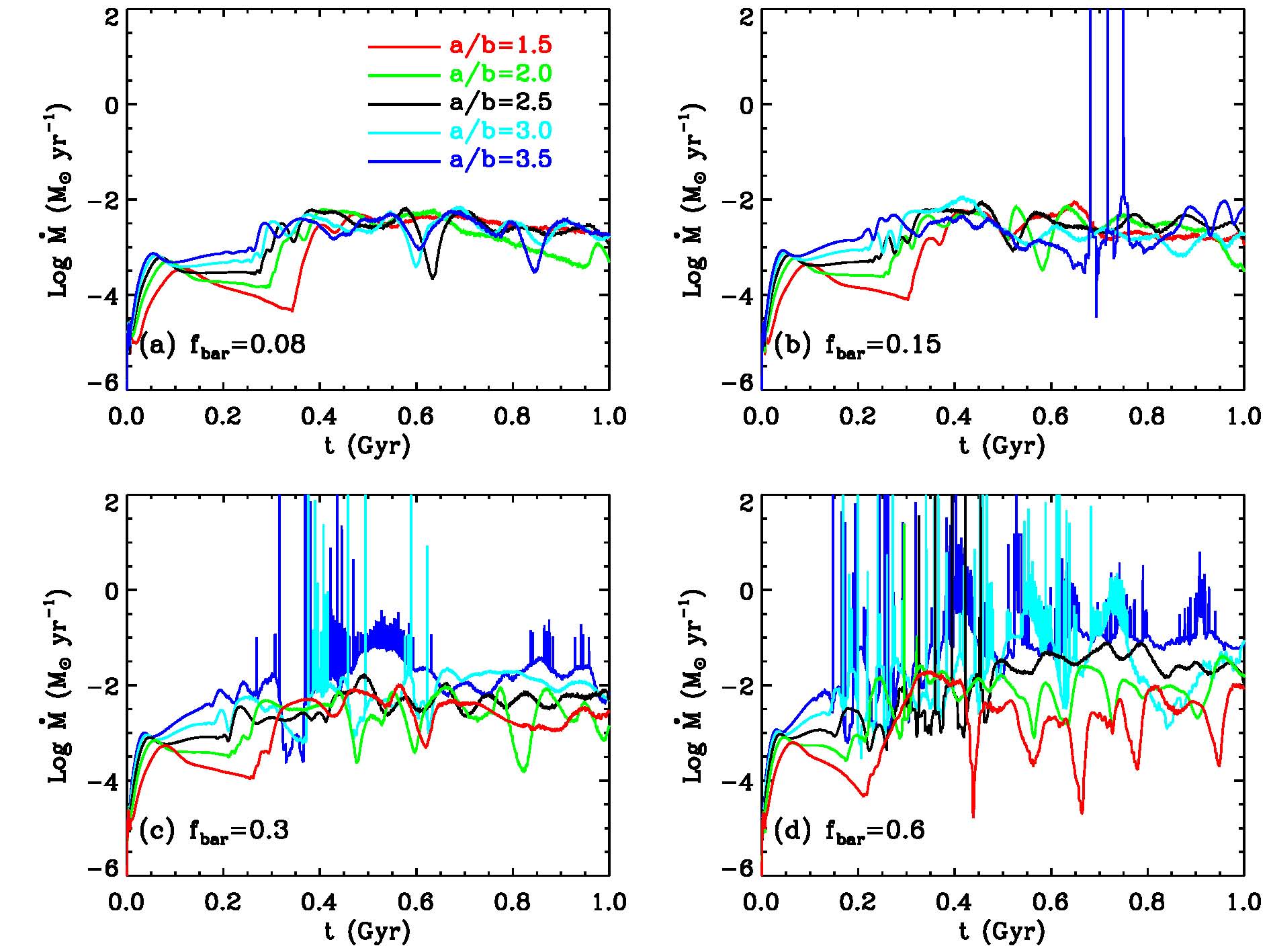}
\caption{ Same as Figure \ref{fig:mdot_no}, but for self-gravitating
models. \label{fig:mdot_sg}}
\end{figure*}

A galactic bar has often been considered to be a powerful means to
transport the interstellar gas at $\sim$kpc scales all the way to
the galaxy center, fueling AGN. To check the viability of this idea
using our models, Figures \ref{fig:mdot_no} and \ref{fig:mdot_sg},
for non-self-gravitating and self-gravitating models, respectively,
plot the mass inflow rates $\Mdot$ through the inner boundary as
functions of time. Table \ref{tbl:mdot} gives the time-averaged
values of $\Mdot$ and the standard deviations $\Delta \Mdot$ over
$t=0.1-1.0\Gyr$ for all models.  One should be cautious in relating
$\Mdot$ to the accretion rate to a central BH, since the inflowing
gas through the inner boundary may change its orbit due possibly to
thermal and radiation pressures, gravity, etc.\ before reaching a BH
and then come out of the inner boundary, or may be lost to star
formation in a circumnuclear disk surrounding a BH (e.g.,
\citealt{kaw08}), which are not captured in our simulations.
Therefore, $\Mdot$ calculated from the current models can be
considered as upper limits to the real accretion rates to a central
BH.

In non-self-gravitating models with a well-defined $\xtwo$-type
ring, $\Mdot$ is quite small at $\sim10^{-4}-10^{-3}\Aunit$ at early
time before the dust-lane shocks fully develop.  It then increases
as the gas flows inward along the dust lanes, but only to
$\sim10^{-3}$--$10^{-2}\Aunit$ unless the bar is strongly elongated
with $\R\simgt3.0$.  These relatively small values of $\Mdot$ result
from the fact that the majority of the inflowing gas from dust lanes
is trapped in a nuclear ring that is located away from the center.
Late time evolution of $\Mdot$ depends on the size of a ring. In
models with $\Rring\simgt0.5\kpc$, $\Mdot$ remains below
$\sim10^{-2}\Aunit$.  In Model M30R25N, on the other hand, the ring
decays with time to have $\Rring\sim0.4\kpc$ at $t=0.6\Gyr$.  At
this small radius, thermal pressure is efficient to perturb
$\xtwo$-like gas orbits into eccentric shapes, increasing $\Mdot$ to
$\sim1\Aunit$ at $t\simgt 0.8\Gyr$.

For models whose central regions are dominated by $\xone$-orbits, on
the other hand, $\Mdot$ increases dramatically to above
$\sim1\Aunit$, and sometimes as large as $\sim10\Aunit$, as an
inclined ring becomes smaller and more eccentric.  When the short
axis of an eccentric ring touches the inner boundary, the ring
material flows in directly through it. This in turn causes the ring
to decay rapidly in time, making $\Mdot$ drop to $\sim
10^{-2}-10^{-1}\Aunit$.  In Model M60R30N, gaseous blobs located in
between the ring and the bar ends gradually move in toward the
center and orbit along $\xone$-orbits in the vicinity of the hole,
which increases $\Mdot$ again to $\sim1\Aunit$ at $t=0.8\Gyr$.

In self-gravitating models with $\Qb\simlt0.2$, $\Mdot$ is slightly
smaller, owing to a larger nuclear ring, than that in the
non-self-gravitating counterpart. In models with $\Qb\simgt0.2$,
however, self-gravity makes the rings unstable, producing small
dense clumps.  While these clumps move along the rings, they
interact with each other and sometimes plunge into the central hole
directly, increasing $\Mdot$ instantaneously. In Model M15R35G with
$\Qb=0.21$, such events occurring three times over 1 Gyr lead to
$\Mdot> 10^2\Aunit$. In other models with larger $\Qb$ such as
Models M30R30G and M60R35G, direct accretion of dense clumps occur
much more frequently, making the $\Mdot$ curves \emph{vs.} time
highly intermittent.

\section{Summary and Discussion}\label{sec:sum}

We have presented the results of high-resolution hydrodynamic
simulations on the formation and evolution of gaseous substructures
in barred galaxies with varying bar strength. We initially consider
an infinitesimally-thin, isothermal, unmagnetized gas disk with
uniform surface density embedded in the external gravitational
potential. We run both non-self-gravitating and self-gravitating
models, but the effects of star formation and feedback are not
considered in the present work.  In order to focus on the effects of
the bar parameters, we fix the gas sound speed to $\cs=10\kms$ and
the BH mass to $\MBH=4\times10^7\Msun$ that affects the rotation
curve near the galaxy center, and vary two parameters: the bar mass
measured by $\fbar$ (Eq.\ [\ref{eq:fbar}]) relative to the
spheroidal component and its aspect ratio $\R$ (see Tables
\ref{tbl:model_no} and \ref{tbl:model_sg} for model parameters).

In what follows, we summarize the main results of the present work
and discuss them in comparison with observations.

\begin{figure}
\hspace{0.2cm}\includegraphics[angle=0, width=8.5cm]{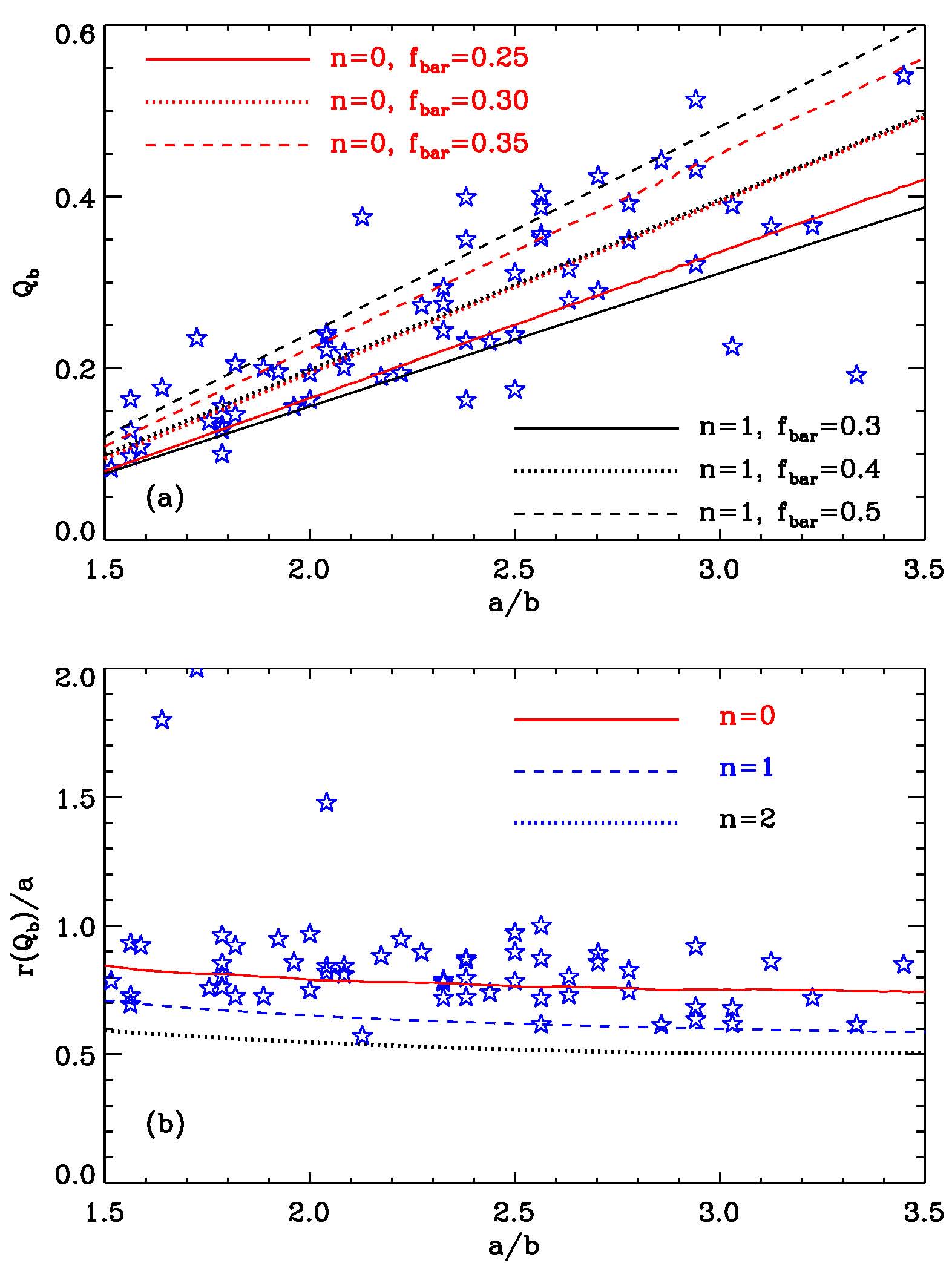}
\caption{ Relationships (a) between $\Qb$ and $\R=a/b$ and (b)
between $\RQ$ and $\R$ from our galaxy models (various lines) with a
Ferrers prolate bar (Eqs.\ [\ref{eq:Qbnum}] and [\ref{eq:Rmaxnum}])
in comparison with observational results (star symbols) of
\citet{com10}. The observed bars are best represented by
$\fbar=0.25$--$0.5$ and $n\leq1$. \label{fig:comp_bar}}
\end{figure}

\emph{1. Bar Strength Parameter} -- We measure the bar strength
using the dimensionless parameter $\Qb$ (Eq.\ [\ref{eq:Qb}]) as is
usually done in observational studies. For our galaxy models with a
Ferrers prolate bar with index $n$, we calculate $\Qb$ as well as
the radius $\RQ$ where the maximum bar torque occurs, and provide
the fitting formulae (Eqs.\ [\ref{eq:Qbnum}] and [\ref{eq:Rmaxnum}])
as functions of $\fbar$ and $\R$.  While $\Qb$ is linearly
proportional to $\R$ and almost linearly to $\fbar$, $\RQ$ is a
weakly decreasing function of $\R$ and independent of $\fbar$. Both
$\Qb$ and $\RQ$ become smaller for a more centrally-concentrated
bar.

Having found the dependence of $\Qb$ and $\RQ$ on the other bar
parameters in our galaxy models, it is interesting to apply our
results to observed barred galaxies.  Recently, \citet{com10}
measured $\Qb$, $\RQ$, $\R$ (or, equivalently, the bar ellipticity),
and the bar semimajor axis $a$ for a sample of nearby galaxies that
contain nuclear rings, which is by far the most complete sample.
Figure \ref{fig:comp_bar}a plots as star symbols the empirical
relation between $\Qb$ and $\R$ from \citet{com10} for galaxies with
$1.5\leq\R\leq3.5$. Also plotted are equation (\ref{eq:Qbnum}) for
various values of $n$ and $\fbar$.   Note that the trend of $\Qb$
becoming larger for larger $\R$ in the observational estimates is
entirely consistent with the results of our galaxy models.

When approximating the observed bars using Ferrers prolate
spheroids, the observational results for $\Qb$ \emph{vs.} $\R$ can
be best described by the bar mass fraction of $\fbar=0.3$--$0.5$ for
$n=1$ inhomogeneous bars and $\fbar=0.25$--$0.35$ for $n=0$
homogeneous bars. The observed relation between $\RQ/a$ and $\R$
shown in Figure \ref{fig:comp_bar}b appears better explained by the
$n=0$ homogeneous bars than inhomogeneous bars.\footnote{Galaxies
with $\RQ/a>1$ have non-axisymmetric torques dominated by outer
spiral arms rather than by bars \citep{com10}.} We note that many
uncertainties surround the observational determinations of $\Qb$ and
$\RQ$ as they rely sensitively on the bulge subtraction, assumptions
on the disk scale height and orientation angle, etc., which are
quite uncertain (e.g., \citealt{lau04,lau06,but06}). Given that
observational errors involved in the $\Qb$ (also likely in $\RQ$)
determinations are typically $\sim20\%$ (e.g., \citealt{com09}), the
comparison shown in Figure \ref{fig:comp_bar} suggests that bars in
real galaxies are most likely to have a mass fraction
$\fbar=0.25$--$0.5$ of the spheroidal component, and unlikely to be
more centrally concentrated than the $n=1$ case.

\begin{figure}
\hspace{0.2cm}\includegraphics[angle=0, width=8.5cm]{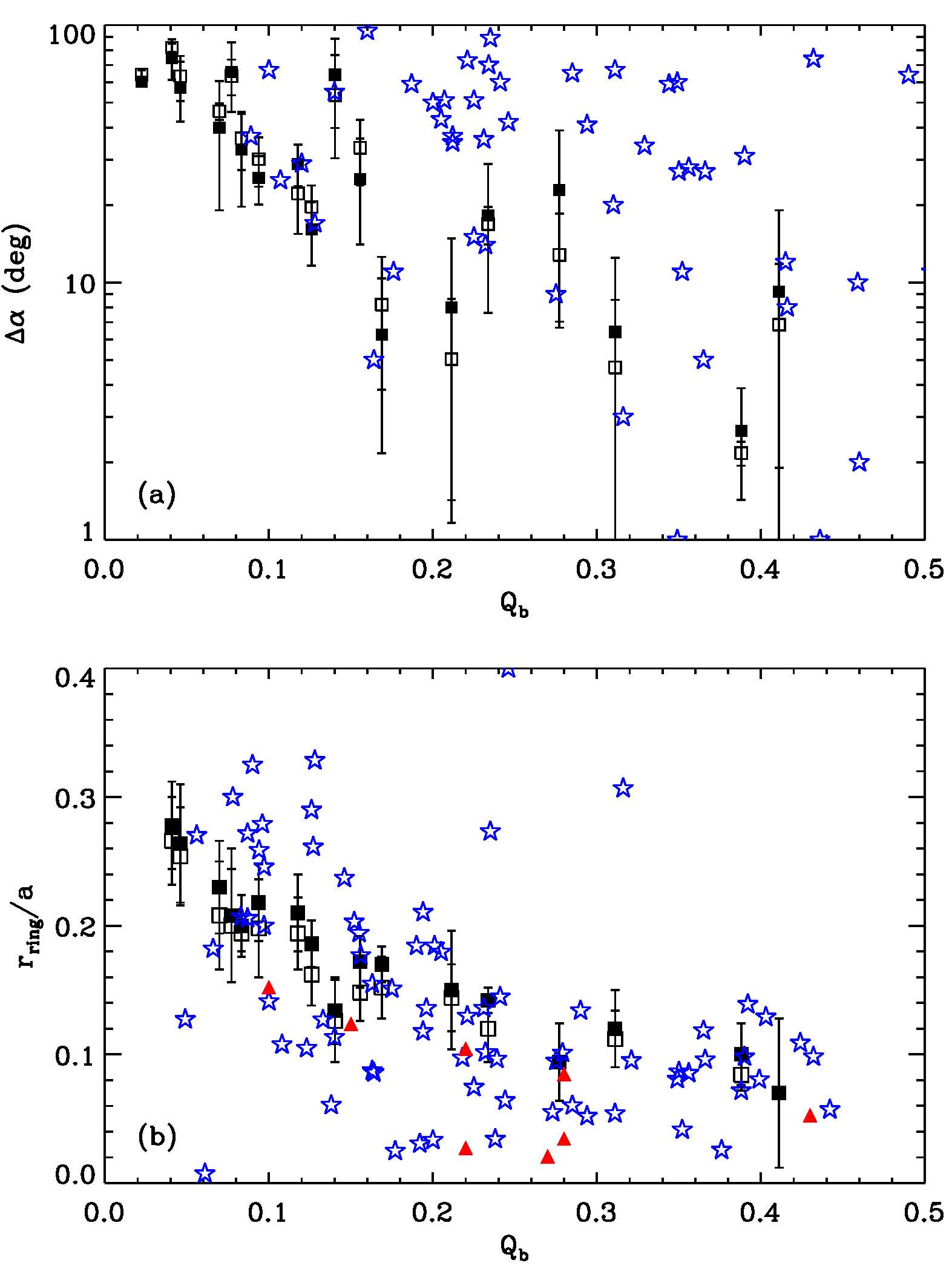}
\caption{Comparison of our numerical results (squares with
errorbars) with the observational measurements for the relations (a)
between the dust-lane curvature $\Dalp$ and $\Qb$ and (b) between
the ring radius $\Rring$ and $\Qb$.  Open and filled squares are for
non-self-gravitating and self-gravitating models, respectively. In
(a), star symbols are adopted from \citet{com09}, while star symbols
and filled triangles in (b) are from \citet{com10} and
\citet{maz11}, respectively. \label{fig:comp_ring}}
\end{figure}

\emph{2. Dust Lanes} -- The imposed non-axisymmetric bar potential
readily induces dust-lane shocks across which gas in rotation about
the galaxy center loses angular momentum significantly and falls
radially inward. Dust lanes in a quasi-steady state approximately
follow one of $\xone$-orbits aligned parallel to the bar major axis.
The curvature $\Dalp$ of dust lanes in our models depends primarily
on $\Qb$ in such a way that they tend to be more curved under a
weaker bar. This results from the facts that dust lanes are closer
to the bar major axis as $\Qb$ increases and that inner
$\xone$-orbits are more eccentric than outer ones (see Fig.\
\ref{fig:orb_comp}). It also depends, albeit less sensitively, on
$\R$ in that a more elongated bar has more straight dust lanes (Eq.\
[\ref{eq:dalpfit}]), since $\xone$-orbits are rounder with smaller
$\R$ when $\Qb$ is fixed. Dust lanes are not much affected by
self-gravity since they have a Toomre stability parameter greater
than unity and are characterized by strong velocity shear.

As mentioned earlier, that dust lanes are more straight under a
stronger bar potential was first theoretically predicted by
\citet{ath92b} and later confirmed empirically by \citet{kna02} and
\citet{com09}. In particular, \citet{com09} measured the dust-lane
curvatures in a sample of 55 barred galaxies that contain clear dust
lanes in the SDSS DR7 or NED images, and studied a relationship
between $\Dalp$ and $\Qb$. Figure \ref{fig:comp_ring}a reproduces
their observational results as star symbols in comparison with our
numerical results shown as open and filled squares with errorbars
for non-self-gravitating and self-gravitating models, respectively.
Note that $\Dalp$ decreases with increasing $\Qb$ in both numerical
and observational results, although $\Dalp$ in the simulations
corresponds roughly to the lower envelope of the observational
results. Note also that our numerical results are unable to
reproduce the scatter seen in the observations. These quantitative
differences of $\Dalp$ between our numerical work and the work of
\citet{com09} are likely due to the differences in the spatial
ranges of dust lanes where $\Dalp$ is measured. \citet{com09}
considered a constant-curvature range that varies from galaxy to
galaxy, while we fix the position angles of both ends of the range.
Also, the fact that the parameter space covered by our models is
very limited (i.e., fixed sound speed, $n=1$ Ferrers bar potential,
no magnetic field, etc.) may also be partly responsible for the
differences between our numerical results and the observations.

\citet{com09} further noted that there is a large spread of $\Dalp$
for given $\Qb$. To find the origin of the spread, they ran a large
number of numerical simulations with differing bar and bulge
parameters, and found that the spread in $\Dalp$ can be reduced if
the aspect ratio $\R$ is considered together with $\Qb$ when fitting
the curvatures.   We similarly found that a linear combination of
$\Qb$ and $\R$ provides a better fit than $\Qb$ alone, although
$\Dalp$ decreases with increasing $\R$ in our fit (Eq.\
[\ref{eq:dalpfit}]), while $\Dalp$ is an increasing function of $\R$
in their fit (Eq.\ [3] of \citealt{com09}). This discrepancy is
again thought of as arising from the differences in the ranges of
dust lanes where $\Dalp$ is measured and from the limited range of
the parameter space in our models, as mentioned above.

\emph{3. Nuclear Rings: Size} -- The shocked gas moving in toward
the central regions along the dust lanes has so large a speed that
the bar torque cannot stop its motion across the ILR. The inflowing
gas keeps moving in and eventually forms a nuclear ring at the
location where the centrifugal force balances the external
gravitational force. The mean radius of a ring in our models is
generally smaller than the ILR location, and decreases
systematically with increasing $\Qb$ (Eqs.\ [\ref{eq:Rring_a}] and
[\ref{eq:Rring_ILR}]). By making the total gravitatioal potential
deeper, self-gravity makes the ring larger in size by $\sim5-20\%$,
with larger values corresponding to lager $\Qb$. Combined with the
results of Paper I that showed that the ring position is insensitive
to the mass of a central BH (and thus to the number and locations of
ILRs), this clearly evidences that the ring position is not
determined by the resonant interactions of gas with the underlying
gravitational potential but rather by the amount of angular momentum
loss at the dust-lane shocks.

\citet{com10} also presented the sizes of nuclear rings in their
sample of barred spiral galaxies.  Figure \ref{fig:comp_ring}b
compares their results (star symbols) with our simulation outcomes
(squares with errorbars) on the $\Rring/a$--$\Qb$ plane. Both
observational and numerical results show that $\Rring/a$ becomes
smaller with increasing $\Qb$, indicating that stronger bars can
possess smaller rings. For $\Qb\simlt0.15$, the agreement between
observational and numerical results is quite good. For
$\Qb\simgt0.15$, on the other hand, $\Rring/a$ in our models
corresponds roughly to the upper envelope of the observational
results for given $\Qb$. This is presumably because our numerical
models are unmagnetized: inclusion of magnetic fields efficiently
removes angular momentum further at dust-lane shocks, which makes
the ring size smaller by a factor of $\sim2$ when magnetic fields
have an equipartition strength with the thermal energy \citep{ks12}.
A larger effective sound speed can additionally make the rings
smaller (Paper I).

More recently, \citet{maz11} measured the ring radii for a sample of
13 barred/unbarred galaxies that contain star-forming nuclear rings:
their results for 8 galaxies whose bar sizes are given in
\citet{com10} are plotted in Figure \ref{fig:comp_ring}b as
triangles. \citet{maz11} argued that the ring size is well
correlated with the compactness $\C\equiv v_0^2/r_t$, where $r_t$ is
the turnover radius of the rotation curve that has the velocity
$v_0$ at the flat part, such that more compact (with smaller
turnover radius) galaxies have a smaller ring. In our galaxy models,
however, $r_t\sim1\kpc$ and $\C\sim 4\times10^4 \;(\rm km\;s^{-1})^2
\kpc^{-1}$, insensitive to $\Qb$ and $\R$, while the variation of
the ring size with $\Qb$ is about by a factor of $\sim2$--$3$.
Since the sample galaxies in \citet{maz11} also exhibit a positive
correlation between $\C$ and $\Qb$, the negative correlation between
$\C$ and $\Rring$ in their results may simply be a reflection of a
more intrinsic negative correlation between $\Rring$ and $\Qb$.  Of
course, our current models with a limited range of the parameters
cannot address the effects of $\C$ on the ring size.

\emph{4. Nuclear Rings: Shape} -- Not all the rings have a
conventional shape similar to $\xtwo$-orbits in our models unless
self-gravity is included. In non-self-gravitating models with $\fbar
= 0.6$ and $\R\geq 2.0$,  the inflowing gas cannot settle on
$\xtwo$-orbits since they have too small kinetic energy.  The ring
material instead takes on an inclined orbit in between $\xone$- and
$\xtwo$-families. An inclined ring in these models becomes smaller
and more eccentric as low angular-momentum gas is added from
outside.  It loses much of its mass through the inner boundary when
its short axis moves close to the center.  Due to the bar torque,
the inclined ring precesses slowly to align its long axis parallel
to the bar major axis, eventually forming an $\xone$-type ring.
Strong self-gravity of an inclined ring provides additional
non-axisymmetric torque that prevents the precession of an inclined
ring, leading instead to an $\xtwo$-type ring in self-gravitating
models.

When $\fbar \leq 0.3$ or $\R\leq 1.5$, on the other hand, the
$\xtwo$-family of closed orbits near the center have sufficiently
large kinetic energy that the inflowing gas along the dust lanes can
transit easily to one of them in both non-self-gravitating and
self-gravitating models. These rings are eccentric with an
ellipticity of $\ering\sim0.2$--$0.3$ insensitive to $\Qb$, which is
within the range of the observed ring ellipticities, $\sim0$--$0.4$,
reported by \citet{com10} and \citet{maz11}. Again, magnetic fields
are expected to circularize nuclear rings \citep{ks12}.

\emph{5. Nuclear Spirals} -- Well-defined twin-armed nuclear spirals
grow only in models in which the nuclear ring is of the $\xtwo$-type
and sufficiently large in size: they would otherwise be destroyed by
the ring material on highly eccentric orbits.  Even in models with
an $\xtwo$-type ring, nuclear spirals are absent if the ring is so
small to limit their spatial extent. All nuclear spirals in our
models are trailing and logarithmic in shape. While the shapes of
dust lanes and nuclear rings do not change much with time after the
potential is fully turned on, nuclear spirals are not stationary
over the course of the entire evolution. They initially start out as
being tightly wound and weak, and then gradually unwind and become
stronger until turning into shocks. This unwinding and growth of
nuclear spirals appears to be a generic property of nonlinear waves
that become more nonlinear as they propagate inward \citep{lee99}.
The unwinding rate is lower in self-gravitating models than in
non-self-gravitating models. Since nuclear spirals grow and unwind
faster as $\Qb$ increases, the probability of having more
tightly-wound and weaker spirals is larger for galaxies with a
weaker bar torque. This is consistent with the observational results
of \citet{pee06} who found that tightly wound spirals are found
primarily in weakly barred galaxies, while loosely wound spirals are
more common in strongly barred galaxies (see also
\citealt{mar03a,mar03b}). \citet{pee06} also found that grand-design
nuclear spirals in strongly barred galaxies does not extend all the
way into the nucleus, which is consistent with our numerical results
that show that nuclear spirals cease to exist by turning into
shocks, and this happens earlier in higher-$\Qb$ models.

\emph{6. Mass Inflow Rates} -- In our models, the mass inflow rate
to through the inner boundary is found to be
$\Mdot\sim10^{-3}-10^{-2}\Aunit$ for models with $\Qb\simlt0.2$
regardless of the presence of self-gravity. Without self-gravity,
models with $\Qb\simgt0.2$ but with an $\xtwo$-type ring still have
$\Mdot < 10^{-2}\Aunit$, since most of the inflowing gas is trapped
in the ring.  If the inflowing gas moves all the way to a central
BH, these values of $\Mdot$ correspond to the Eddington ratio
$\lambda\equiv L_{\rm bol}/L_{\rm Edd} = 4.5\times10^{-2}
(\Mdot/10^{-2}\Aunit)(\MBH/10^7\Msun)^{-1} \sim 10^{-3}$--$10^{-2}$
(Paper I), potentially explaining low-luminosity Seyfert 1 AGNs.
(e.g., \citealt{ho08}). Here, $L_{\rm bol}$ and $L_{\rm Edd}$ denote
the bolometric and Eddington luminosities of an AGN, respectively,
and 10\% of the mass-to-energy conversion efficiency of the accreted
material is assumed. Some of our numerical models exhibit
unrealistically large mass inflow rates. Non-self-gravitating models
in which the central regions are dominated by the gas on
$\xone$-orbits are found to have $\Mdot \simgt 1\Aunit$ and
sometimes as large as $\sim10\Aunit$ when the ring gas on eccentric
orbits is accreted directly to the inner boundary. In
self-gravitating models with $\Qb\simgt0.2$, on the other hand,
rings are unstable to form high-density clumps with mass
$\sim10^6-10^7\Msun$.  These clumps sometimes plunge into the
central hole, causing the mass inflow rate to fluctuate with large
amplitudes.

We finally remark some caveats associated with $\Mdot$ obtained in
our simulations and in interpreting it as a mass accretion rate to a
central BH. First, $\Mdot$ from self-gravitating models are
definitely more realistic that that from non-self0gravitating
models. Still, Models M60R30G and M60R35G can not be applied to real
galaxies since they do not posses well-defined nuclear rings: the
bar in these models is perhaps too massive or too elongated, or the
Ferrers prolate bar is not a good representation of realistic bars.
Second, as mentioned earlier, $\Mdot$ measured is the rate of gas
mass that goes in through the inner boundary.  This is likely to be
an upper limit to the real accretion rate to the BH since some of
the inflowing mass changes its orbit before reaching the BH and can
possibly come out of the inner boundary. Third, high-density clumps
produced by gravitational instability of nuclear rings would undergo
star formation, reducing gas content in the rings. Ensuing feedback
would destroy them, so that the accretion of dense clumps in real
situations would much less frequent than in our simulations. Fourth,
a circumnuclear disk with starburst activities can make the
accretion rate to the BH much smaller than $\Mdot$ (e.g,
\citealt{dav07,wat08,kaw08}). Fifth, while magnetic fields are known
to enhance $\Mdot$ considerably \citep{ks12}, they would suppress or
reduce gravitational instability of nuclear rings, which tends to
reduce $\Mdot$. In order to properly evaluate the mass accretion
rates, therefore, it is required to run more realistic models of
barred galaxies including star formation, feedback, magnetic fields,
and other physical processes that affect gaseous features and gas
inflows in the nuclear regions.

\acknowledgments We acknowledge helpful and stimulating comments
from S.\ Comer\'on, J.\ Knapen, and P.\ Martini, and a thoughtful
report from the referee.  We are also grateful to R.\ Shetty and E.\
Ostriker for sharing their self-gravity solver based on the Kalnajs
method. This work was supported by the National Research Foundation
of Korea (NRF) grant funded by the Korean government (MEST), No.\
2010-0000712.

\appendix

\section{Bar Strength for A Prolate Ferrers Bar}\label{app:pot}

In this Appendix, we provide analytic expressions for the bar
strength $\Qb$ and the radius $\RQ$ of the maximum bar torque for a
galaxy with an $n=1$ prolate Ferrer bar, whose density distribution
is given by equation (\ref{eq:bar}). The eccentricity of the bar is
$e\equiv (1-b^2/a^2)^{1/2}$. Since the maximum bar torque always
occurs inside the bar, it is sufficient to consider the interior
(i.e., $g\leq 1$) gravitational potential of the bar.

Following the procedures presented by \citet{pfe84} for triaxial
ellipsoids and by \citet{mac00} for prolate spheroids, one can show
that the interior potential of a Ferrers bar with $n=1$ at the $z=0$
plane is given by
\begin{equation}\label{eq:barp}
\Pbar(x,y) = - (\pi/2) G ab^2 \rhobar
(W_{00} - 2W_{01} x^2 - 2W_{10}y^2 + W_{02}x^4 +
2W_{11} x^2y^2 +W_{20}y^4 ),
\end{equation}
with the coefficients
\begin{equation}\label{eq:00}
W_{00} = \frac{1}{ae}\ln \left(\frac{1+e}{1-e}\right) = \frac{I}{a},
\end{equation}
\begin{equation}\label{eq:10}
W_{10} =
\frac{2}{a^3e^2}
\left[ \frac{1}{2e}\ln \left(\frac{1+e}{1-e}\right) -1 \right]
=\frac{A_3}{ab^2},
\end{equation}
\begin{equation}\label{eq:01}
W_{01} = \frac{1}{a^3e^2}
\left[ \frac{1}{1-e^2} - \frac{1}{2e}
\ln \left(\frac{1+e}{1-e}\right) \right]
=\frac{A_1}{ab^2},
\end{equation}
\begin{equation}
W_{11} = (W_{01}-W_{10})/(a^2e^2),
\end{equation}
\begin{equation}
W_{20} = \frac{2}{3}\left(\frac{1}{a^5(1-e^2)} - W_{11} \right),
\end{equation}
and
\begin{equation}
W_{02} = \frac{1}{4}\left(\frac{2}{a^5(1-e^2)^2} - W_{11}\right).
\end{equation}
Note that $I$, $A_1$, and $A_3$ in equations (\ref{eq:00}) -- (\ref{eq:01})
are identical to the dimensionless coefficients for the gravitational
potential of uniform spheroids tabulated in Table 2.1 of
\citet{bin08}.

The gravitational force in the azimuthal direction due to the
$n=1$ prolate is then given by
\begin{equation}
F_T = -\frac{1}{r}\frac{\partial\Pbar}{\partial\phi} =
- 2\pi G ab^2\rhobar xy(C_x x^2 + C_0 + C_y y^2 )/r,
\end{equation}
where
\begin{eqnarray}
C_x &= W_{02}-W_{11}, \\
C_0 &= W_{10}-W_{01},  \\
C_y &= W_{11}-W_{20}.
\end{eqnarray}
It can be easily verified that $C_x,C_y\geq0$ and $C_0\leq0$ for $0\leq e <1$.
In the limit of $e\rightarrow 0$, $C_x=C_y\rightarrow 2e^2/(7a^5)$,
while $C_0\rightarrow -2e^2/(5a^3)$.

It is straightforward to show that the maximum value of $|F_T|$ is
\begin{equation}
F_{T,\rm max} = \frac{4\pi G ab^2\rhobar}{3^{3/2}}
\frac{(-C_0)^{3/2}} {C_x^{1/2} + C_y^{1/2}},
\end{equation}
which occurs at
$x^2 = -C_0/(3C_x^{1/2})(C_x^{1/2} + C_y^{1/2})^{-1}$
and $y^2 = -C_0/(3C_y^{1/2})(C_x^{1/2} + C_y^{1/2})^{-1}$.

For galaxies with a constant rotational velocity $v_0$,
the radial gravitational force is $F_R=v_0^2/r$.  Therefore, the bar
strength parameter defined in equation (\ref{eq:Qb}) becomes
\begin{equation}
\Qb = \frac{(rF_T)_{\rm max}}{v_0^2}
= \frac{\pi G ab^2 \rhobar}{4v_0^2} \frac{C_0^2}{\sqrt{C_xC_y}},
\end{equation}
which is attained at $x^2=-C_0/(4C_x)$ and $y^2=-C_0/(4C_y)$, or
at the radius
\begin{equation}\label{eq:Rmax}
\RQ = \frac{(-C_0)^{1/2}}{2}\left(\frac{1}{C_x}+\frac{1}{C_y}\right)^{1/2}.
\end{equation}
Let $\Mtot(a)=av_0^2/G$ denote the total galaxy mass within $r<a$.
Assuming that the galaxy mass is dominated by the bulge and the bar
inside $r=a$, it then follows that
\begin{equation}\label{eq:Qbana}
\Qb=\frac{15}{32}
\frac{aC_0^2}{\sqrt{C_xC_y}}\fbar.
\end{equation}
Figure \ref{fig:Qb_rmax} plots $\Qb$ and $\RQ$ in equations
(\ref{eq:Qbana}) and (\ref{eq:Rmax})
as dotted lines, which is not much different from the true bar strength
that does not assume flat rotation.

\pagebreak

\end{document}